\documentclass{aa}  
\usepackage{graphicx}
\usepackage[varg]{txfonts}
\usepackage{longtable}
\usepackage{natbib}
\usepackage{url}
\usepackage{color}
\usepackage{multirow,bigdelim}
\usepackage{cases}
\usepackage{rotating}
\usepackage[colorinlistoftodos]{todonotes}
\bibpunct{(}{)}{;}{a}{}{,} 
\newcommand\kms{{\rm\,km\,s^{-1}}}

\usepackage{hyperref} 
\hypersetup{
     breaklinks,
     colorlinks   = true,
     citecolor    = blue,
     urlcolor     = blue,
     linkcolor    = blue
}

\begin{document} 

\title{The HR~1614 moving group is not a dissolving cluster}

\titlerunning{The HR~1614 moving group is not a dissolving cluster} 

\author{
Iryna Kushniruk \inst{1},
Thomas Bensby \inst{1},
Sofia Feltzing \inst{1},
Christian L. Sahlholdt \inst{1},
Diane Feuillet\inst{1},
Luca Casagrande\inst{2}
}

\authorrunning{
I. Kushniruk et al.
}

\institute{
Lund Observatory, Department of Astronomy and Theoretical Physics, Box 43, SE-221\,00 Lund, Sweden
\and
Research School of Astronomy \& Astrophysics, Mount Stromlo Observatory, The Australian National University, ACT 2611, Australia\\
\email{[iryna; tbensby]@astro.lu.se}
}

\date{Received 11 March 2020 / Accepted 11 May 2020}

 
\abstract
{The HR~1614 is an overdensity in velocity space and has for a long time been known as an old ($\sim 2$\,Gyr) and metal-rich ($\rm [Fe/H]\approx+0.2$) nearby moving group that has a dissolving open cluster origin. The existence of such old and metal-rich groups in the solar vicinity is quite unexpected since the vast majority of nearby moving groups are known to be young.
}
{
In the light of new and significantly larger data sets than ever before (astrometric, photometric, and spectroscopic), we aim to re-investigate the properties and origin of the HR~1614 moving group. If the HR~1614 overdensity is a dissolving cluster, its stars should represent a single-age and single-elemental abundance population.
}
{
To identify and characterise the HR~1614 moving group we use astrometric data from {\it Gaia} DR2; distances, extinction, and reddening corrections from the {\tt StarHorse} code; elemental abundances from the GALAH and APOGEE spectroscopic surveys; and photometric metallicities from the SkyMapper survey. Bayesian ages were estimated for the SkyMapper stars. Since the Hercules stream is the closest kinematical structure to the HR~1614 moving group in velocity space and as its origin is believed to be well-understood, we use the Hercules stream for comparison purposes. Stars that are likely to be members of the two groups were selected based on their space velocities.   
}
{
The HR~1614 moving group is located mainly at negative $U$ velocities, does not form an arch of constant energy in the $U-V$ space, and is tilted in $V$. We find that the HR~1614 overdensity is not chemically homogeneous, but that its stars exist at a wide range of metallicities, ages, and elemental abundance ratios. They are essentially similar to what is observed in the Galactic thin and thick discs, a younger population (around 3\,Gyr) that is metal-rich ($-0.2\le$ [Fe/H] $\le0.4$) and alpha-poor. These findings are very similar to what is seen for the Hercules stream, which is believed to have a dynamical origin and consists of regular stars from the Galactic discs. 
}
{
The HR~1614 overdensity has a wide spread in metallicity, [Mg/Fe], and age distributions resembling the general properties of the Galactic disc. It should therefore not be considered a dissolving open cluster, or an accreted population. Based on the kinematic and chemical properties of the HR~1614 overdensity we suggest that it has a complex origin that could be explained by combining several different mechanisms such as resonances with the Galactic bar and spiral structure, phase mixing of dissolving spiral structure, and phase mixing due to an external perturbation.  
}


\maketitle
%

\section{Introduction}

The Milky Way has a complex structure \citep[e.g.][]{_bh16}. By studying how stars move in the Galaxy we can trace the formation history of the Milky Way \citep[e.g.][]{_freeman02}. Stars that share a common motion are usually called kinematic structures, and studying their origin is one of the ways of obtaining more information about the formation and evolution of the Galactic disc. The analysis of data from the {\it Hipparcos} mission \citep{_perryman97} revealed a rich structure of the local velocity distribution \citep[e.g.][]{_dehnen00, _arifyanto06}. Later, an even more complex picture was discovered by the {\it Gaia} mission \citep{_gaia16}. The local velocity field is composed of dozens of kinematic structures that together form arches and ridges in different velocity projections \citep[e.g.][]{_antoja18, _katz18k, _ramos18, _khanna19, _kushniruk19}. The origin of the non-smooth local velocity distribution is directly linked to major formation processes of the Milky Way. For example, kinematic structures can have different origins, including resonances with the Galactic bar \citep[e.g.][]{_dehnen00, _monari18}, resonances with the spiral arms \citep[e.g.][]{_quillen18}, phase mixing as a result of a dynamical interaction with a merging dwarf galaxy \citep[e.g.][]{_minchev09, _antoja18, _laporte19}, phase mixing as a result of transient spiral arms and a perturbation with a dwarf satellite \citep[e.g.][]{_khanna19}, and phase mixing due to transient spiral structure \citep[e.g.][]{_hunt19}. 

Another explanation for some kinematic structures is that open clusters dissolve with time and form moving groups. This idea was first introduced by Olin Eggen \citep{_eggen65}. Since stars in a moving group originate from the same cluster, they share similar chemical composition, ages, and motion. Two well-known moving groups are the Pleiades and the Hyades, which are visible to the naked eye. More contended examples of moving groups are the Arcturus and the HR~1614 moving groups that were discovered in \citet{_eggen71} and \citet{_eggen78}. The groups were later confirmed with {\it Hipparcos} and {\it Gaia} data \citep[e.g.][]{_dehnen98, _feltzing00, _ramos18, _kushniruk19}. For a long time the Arcturus structure was considered a dissolving open cluster, later an accreted stellar population. However, \citet{_kushniruk19} used the {\it Gaia} Data Release 2 (DR2) data and found no signatures of it being a dissolving cluster or an accreted population. Instead, the group was likely caused by phase mixing induced by a merger, as was proposed by \citet{_minchev09}. 

A discovery of a group of stars with mean radial velocity component $U\simeq 0 \kms$ and mean rotational velocity component $V\simeq-60 \kms$ and higher than solar metallicity was first reported in \citet{_eggen78}. The group was named after the star HR~1614\footnote{HR1614 has a {\it Gaia} DR2 ID: {\tt 3211461469444773376. ID in the Henry Draper Catalogue is HD 32147.}} which is one of its member stars. It was proposed that the overdensity is a dissolving old open cluster. Later \citet{_eggen92} estimated that the HR~1614 moving group is about 5\,Gyr old. The origin of the group was re-investigated in \citet{_feltzing00} using {\it Hipparcos} data. Candidate member stars of the group were carefully selected by examining Hertzsprung--Russell diagrams (HR diagram) for different slices of the $U-V$ space velocity distribution.  A slice with younger and more metal-rich stars was found. From dynamical simulations of a disrupting open cluster they could explain the tilt of the group in the $U-V$ space. They also found the group to be about 2\,Gyr old, and that its metallicity is $\rm [Fe/H] \simeq 0.19$. Later, \citet{_desilva07} performed a spectroscopic study of stars that were assigned as members of the group by \citet{_feltzing00} and confirmed that HR\,1614 is a 2 Gyr metal-rich group with [Fe/H] $\geq$ 0.25. It was also found to be chemically homogeneous with a scatter of only about 0.01\,dex in various chemical elements. So, from the literature we know that HR~1614 is a metal-rich, 2 Gyr moving group that is scattered around the Sun.

When searching for the Arcturus stream \citet{_kushniruk19} were able to clearly identify the HR~1614 moving group in the velocity distribution. The question is whether HR~1614 can still be considered a dissolved open cluster if the much larger and more precise astrometric sample from {\it Gaia} DR2 is used to identify and characterise its properties in more detail. In this paper we aim to do just that, to analyse kinematic, chemical, and photometric properties of the 1614 moving group in order to constrain its origin using data from {\it Gaia} DR2 and spectroscopic surveys such as APOGEE and GALAH. 

The paper is structured in the following way. The data sets used are described in Sect.~\ref{_sec_data}. The selection of candidate member stars of the HR~1614 moving group is described in Sect.~\ref{_sec_analysis}, where we also analyse the HR diagrams, and the metallicity and age distributions of the HR~1614 moving group stars. In Sect.~\ref{_sec_discussion} we discuss the possible origins of HR~1614, and finally our findings are summarised in Sect.~\ref{_sec_conclusion}.  

\section{Data}\label{_sec_data}
In this work we use proper motions, sky positions, and radial velocities from the {\it Gaia} DR2 \citep{_brown18} catalogue, Bayesian extinction corrections and distances from the {\tt StarHorse} code \citep{_anders19}. We cross-match the {\tt StarHorse} and {\it Gaia} DR2 catalogues by {\it Gaia} IDs where the radial velocity is not NULL, where the internal {\tt StarHorse} quality flag {\tt SH\_OUTFLAG} is set to {\tt`00000'}, and where the {\it Gaia} quality flag {\tt SH\_GAIAFLAG} is set to {\tt `000'}. These are recommended quality flags that are explained in detail in \citet{_anders19}. In addition, we also cut out stars with {\tt parallax\_over\_error $> 10$} and {\tt visibility\_periods\_used $> 8$} as suggested by \citet{_babasiaus18}. The final data query is as follows:  

\begin{verbatim}
SELECT s.*, g.*
FROM gdr2.gaia_source AS g, 
gdr2_contrib.starhorse AS s 
WHERE g.source_id = s.source_id 
AND g.radial_velocity IS NOT NULL
AND s.SH_OUTFLAG LIKE '00000'
AND s.SH_GAIAFLAG LIKE '000'
AND g.parallax_over_error > 10
AND g.visibility_periods_used > 8
\end{verbatim}

The query listed above gives us a sample of 4\,790\,725 stars. We use the {\it galpy}\footnote{Available at \url{https://github.com/jobovy/galpy}} package \citep{_bovy15} to calculate: 1) space velocities $U$, $V$, and $W$\footnote{$U$ points at the Galactic centre, $V$ in the direction of Galactic rotation, and $W$ towards the Galactic north pole.} corrected for peculiar motion of the Sun with ($U_{\odot}, V_{\odot}, W_{\odot}$) = (11.10, 12.24, 7.25) $\kms$ \citep{_schonrich10} and 2) Galactocentric cylindrical coordinates $R$, $\phi$, and $Z$\footnote{$R$ and $\phi$ point in the opposite direction to $U$ and $V$.} with $R_{\odot}$ = 8.34 kpc \citep{_reid14}, $\phi_{\odot} = 0^{\circ}$, and $Z_{\odot}=14$ pc \citep{_binney97}.

To characterise the groups chemically, we use elemental abundances from both the APOGEE \citep{_majewski17} and GALAH \citep{_desilva15} spectroscopic surveys. APOGEE provides stellar atmospheric parameters and elemental abundances derived from near-IR ($H$-band), high-resolution (R $\sim 23 000$) spectra. We use the abundances from APOGEE DR16 \citep{_apogee_dr16}, which includes stars in both the northern and southern hemispheres. After applying the same cuts for quality assurance as in \citet{_feuillet19} and cross-matching our sample with APOGEE by {\it Gaia} IDs, we have a sample of 126\,690 red giants. GALAH is also a high-resolution (R $\sim 28 000$) spectroscopic survey operating at optical wavelengths in the southern hemisphere that provides stellar atmospheric parameters and elemental abundances. We use GALAH DR2 \citet{_buder18}, setting the flag\_cannon and flag\_x\_fe flags to zero as recommended. Cross-matching GALAH with our set of stars results in a sample of 231\,725 stars. We also use data from the SkyMapper \citep{_casagrande19} survey, which provided photometric metallicities for 907\,893 stars after cross-matching it with our sample of {\it Gaia} IDs. To navigate on HR diagrams presented in this paper, we use PARSEC 1.2S isochrones\footnote{Available at \url{http://stev.oapd.inaf.it/cgi-bin/cmd\_3.0}} re-derived for the {\it Gaia} DR2 photometric system \citep{_maiz18}. 

\section{Analysis}\label{_sec_analysis}


\begin{figure}
   \centering
   \resizebox{\hsize}{!}{
   \includegraphics[viewport = 10  0 430 290, clip]{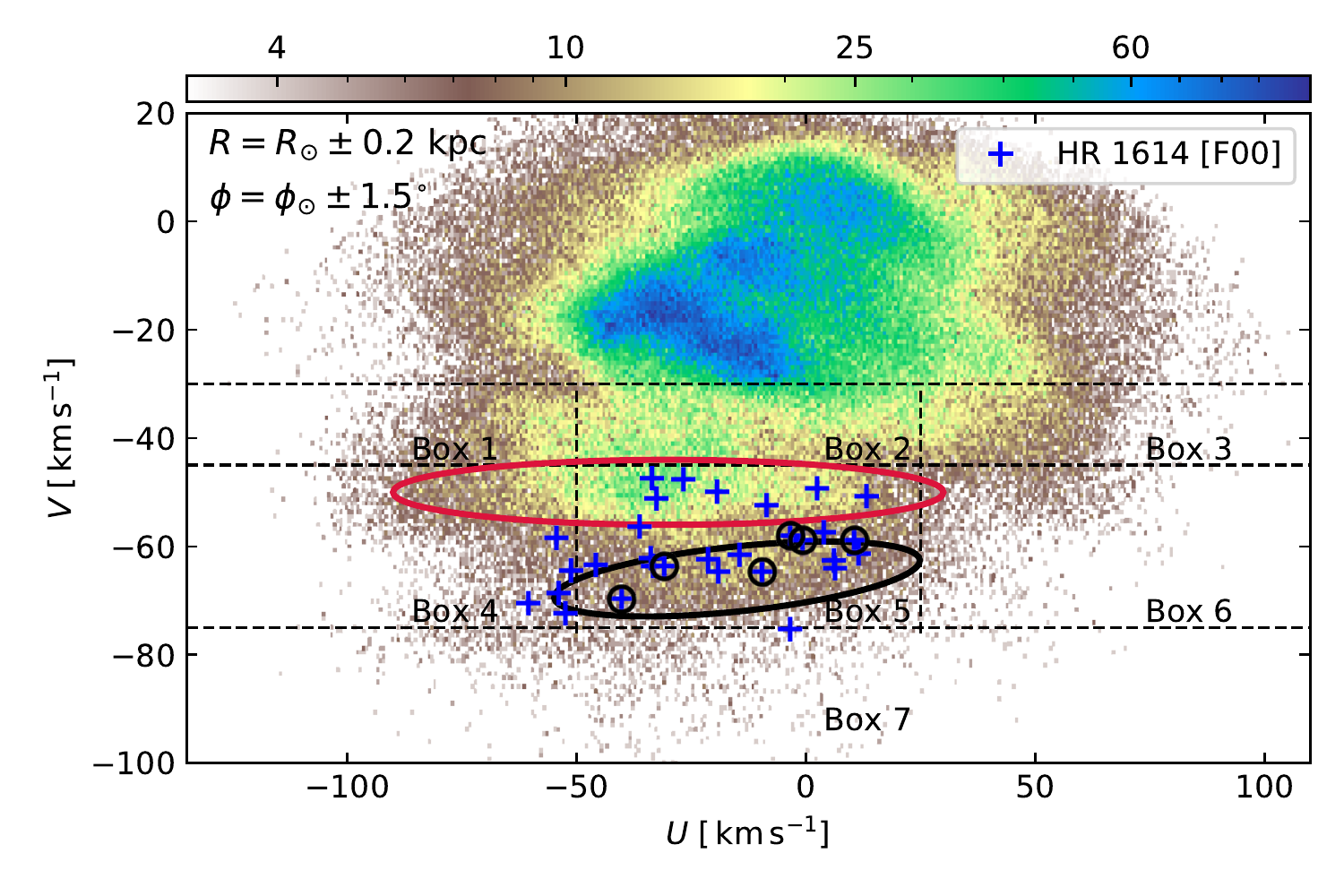}
   }
   \caption{$U-V$ distribution of 581\,190 stars inside a region defined in $R_{\odot} \pm 0.2$ kpc and $\phi_{\odot} \pm 1.5^{\circ}$ (see Sect. \ref{_sec_analysis}). The color scale is proportional to the number of stars, as indicated in the bar at the top. Stars in the HR~1614 moving group from \citet{_feltzing00} are shown as blue crosses and their best-selected candidates as black open circles. We identify stars within the red and black ellipses as members of the Hercules stream and the HR~1614 moving group, respectively (see discussion in Sect.~\ref{_sec_analysis}). Dashed lines and numbers show the same boxes as in \citet{_feltzing00}.
   \label{_fig_uv_ellipses}
   }
\end{figure}

\begin{figure*}
   \centering
   \resizebox{0.8\hsize}{!}{
   \includegraphics[viewport = 0  0 510 480,clip]{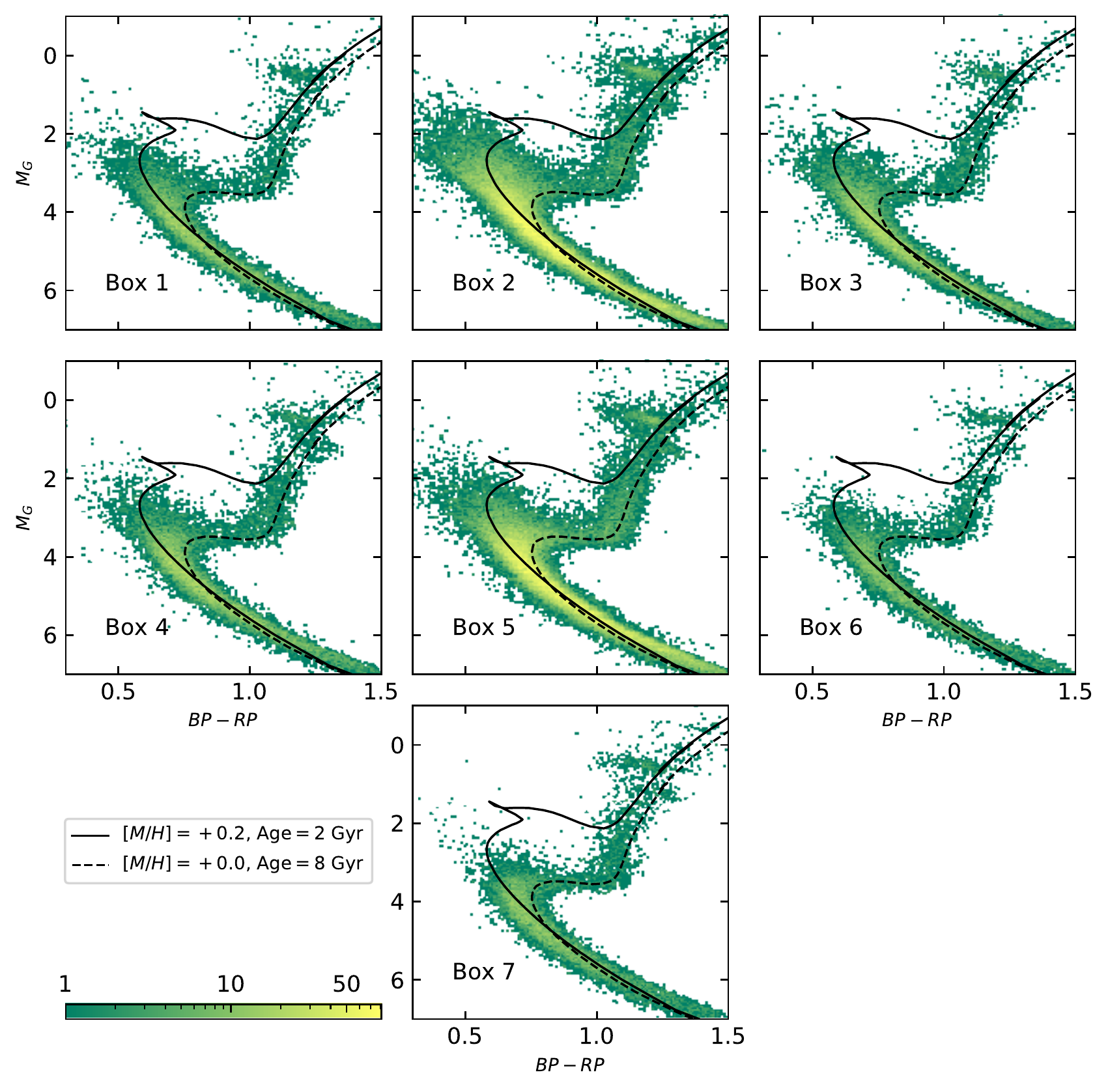}
   }
   \caption{Hertzsprung--Russel diagrams for stars inside the seven boxes as shown in Fig.\ref{_fig_uv_ellipses}. The dashed line is an isochrone for [M/H] $=0$ and 8 Gyr, the solid line shows an isochrone for [M/H] $=+0.2$ and 8 Gyr. The Colour bar shows the number of stars in each bin. 
   \label{_fig_hrd_boxes}
   }
\end{figure*}

\subsection{The old HR~1614 moving group}
The nature of the HR~1614 overdensity was previously investigated by \citet{_feltzing00}. They used Hipparcos data and divided the $U-V$ space into seven smaller boxes. Individual HR diagrams for each box revealed one box with a population that was younger and more metal-rich than the Sun, which was connected to the HR~1614 moving group. 

In this study we repeat the same procedure, but with our much larger stellar sample. Figure~\ref{_fig_uv_ellipses} shows the $U-V$ distribution of 581\,190 stars close to the Sun within a region defined by $R_{\odot} \pm 0.2$ kpc and $\phi_{\odot} \pm 1.5^{\circ}$. This data is sliced into seven boxes, as was done in \citet{_feltzing00}. The boxes are shown with dashed lines in Figure \ref{_fig_uv_ellipses}. Box 5 is where the group is supposed to be located according to \citet{_feltzing00}. With {\it Gaia} DR2/{\tt StarHorse} we clearly see that box 5 contains two velocity overdensities, whereas in \citet{_feltzing00} two separate overdensities were not observed (probably due to a significantly smaller stellar sample). The inclined overdensity in box 5 in Fig.~\ref{_fig_uv_ellipses} is what we link to the HR~1614 moving group (marked by a black ellipse). The horizontal overdensity in box 5 is known as the Hercules stream (marked by a red ellipse). The identification of the groups is based on results of the wavelet transform performed in \citet{_kushniruk19}.

The Hercules stream has been studied in many works: it is likely caused by resonances with the bar \citep[e.g.][]{_dehnen00, _bensby07, _wegg15}, and thus, will be used as a benchmark group in this study. Due to its resonant origin, the Hercules stream mainly consists of stars from the Galactic disk. Fig.~\ref{_fig_uv_ellipses} show stars that are potential members of the HR 1614 group taken from Table 1 in \citet{_feltzing00}. These are the same stars as the ones shown as filled black circles in Table 1 in \citet{_feltzing00}. Interestingly, although \citet{_feltzing00} did not see HR~1614 as a separate overdensity in their box 5, the circled crosses that show stars from their paper and our black ellipse are tilted at similar angles.

Figure~\ref{_fig_hrd_boxes} shows individual HR diagrams for boxes 1-7 using extinction corrected colours and absolute magnitudes from the {\tt StarHorse} catalogue. Two sets of isochrones are shown: one with an age of 2\,Gyr and metallicity $\rm [M/H]=+0.2$ and one with an age of 8\,Gyr and $\rm [M/H]=+0.0$. It is clear that none of the HR diagrams can be fitted with only one isochrone. This probably means that the stars in all boxes represent an underlying population of stars with a wide range in metallicity and/or age. Similarly to what was seen in \citet{_feltzing00}, boxes 2 and 5 appear to contain younger stars compared to the other boxes. These are also the boxes with the largest number of stars. As was shown in Fig.~\ref{_fig_uv_ellipses}, in our larger data set we clearly see that box 5 consists of two different groups, Hercules and HR~1614, that were not observed as separate overdensities with the Hipparcos data \citep{_feltzing00}.

\begin{figure*}
   \centering
   \resizebox{\hsize}{!}{
   \includegraphics[viewport = 0  140 920 770,clip]{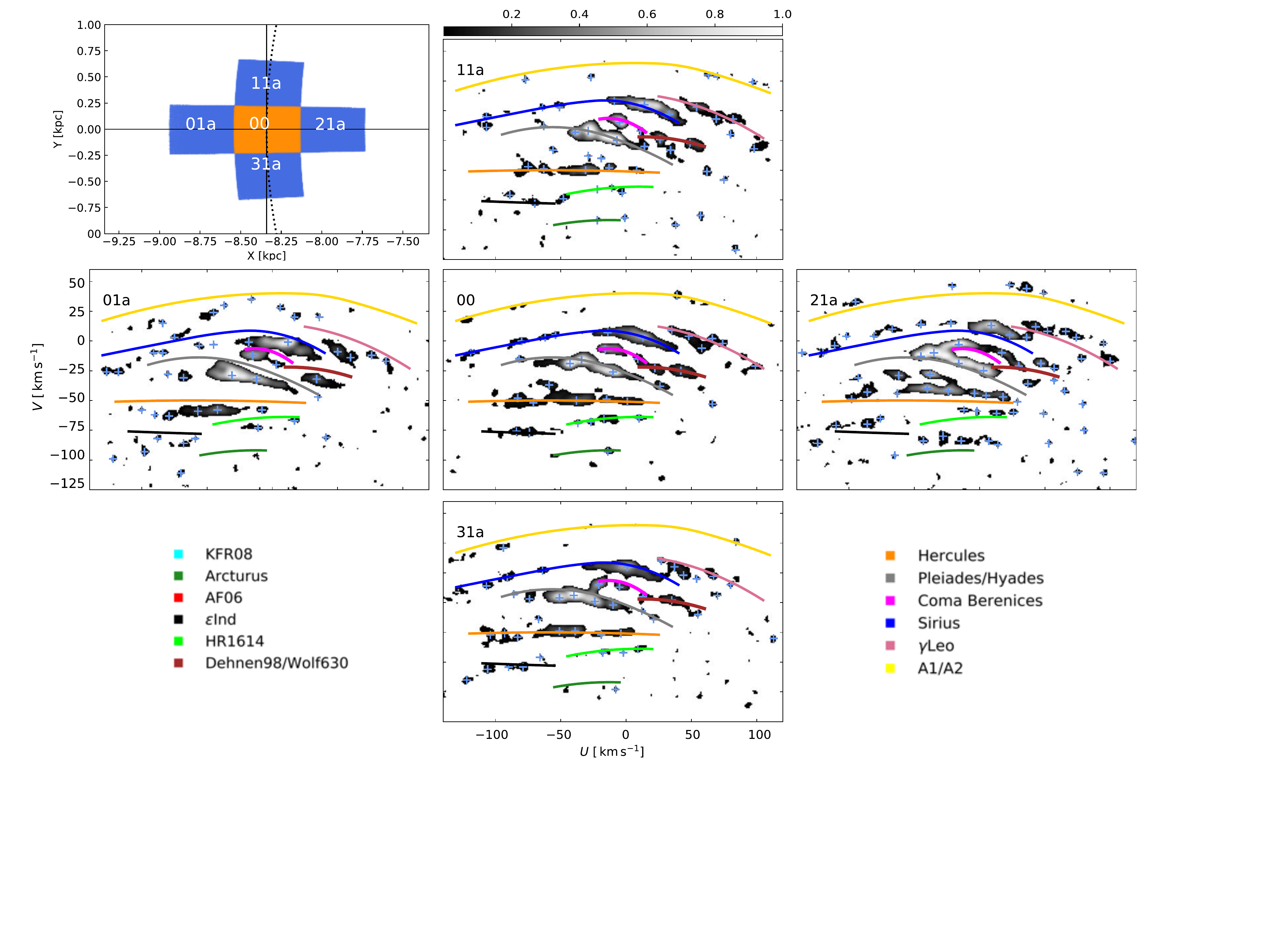}
   }
   \caption{Top left: Location of five regions (00, 01a, 11a, 21a, and 31a) in the $X$ -- $Y$ space explored in this work. Other plots: Wavelet transform maps that show location of the kinematic overdensities in the $U$ -- $V$ space. Their centres are shown with blue crosses. Lines connect eponymous overdensities into structures. The names of the structures are listed in the legend. The colour bar shows the normalised wavelet coefficients in each bin. 
   \label{_fig_uv}
   }
\end{figure*}

\subsection{The new HR~1614 moving group}

A more robust way to select candidate member stars in the HR~1614 moving group is to apply the results of the wavelet transform analysis presented in \citet{_kushniruk19}. They analysed the {\it Gaia} DR2 radial velocity sample in the $U-V$ space by individually exploring smaller sub-samples at different $R$ and $\phi$. Their analysis is based on the discrete wavelet transform, an algorithm that decomposes data into a set of wavelet coefficients. These coefficients contain information about the location and significance of velocity overdensities present in the data. More details on the wavelet transform can be found in \citet{starck_astronomical_2002}, and about the procedure in \citet{_kushniruk17} and \citet{_kushniruk19}. 

The locations of the sub-samples (regions 00, 01a, 11a, 21a, and 31a) from \citet{_kushniruk19} in Cartesian Galactic $X$ and $Y$ coordinate system are shown in the top left plot of Fig.~\ref{_fig_uv}. The other plots in Fig.~\ref{_fig_uv} show wavelet transform maps for the corresponding regions. The HR~1614 moving group is observed as four overdensities with mean $U=-18 \kms$ and $V=-65 \kms$ in region 00. The group is also visible in regions 11a and 31a at the same location and is slightly shifted downwards in $V$ in region 01a and shifted upwards in region 21a. The shift occurs due to the location of the regions at different Galacticentric radii. 

The $U-V$ distribution of stars in Fig.~\ref{_fig_uv_ellipses} is shown for region 00. A black ellipse drawn around the HR~1614 moving group is centred at $U=-18 \kms$ and $V=-65 \kms$, and a red ellipse around the Hercules stream is centred at $U=-37 \kms$ and $V=-49 \kms$. The positions of the two groups were taken from \citet{_kushniruk19}, and the height and width of the ellipse were arbitrarily selected to fit the $U-V$ distribution. 

\begin{figure*}
   \centering
   \resizebox{0.7\hsize}{!}{
   \includegraphics[viewport = 0  0 380 200,clip]{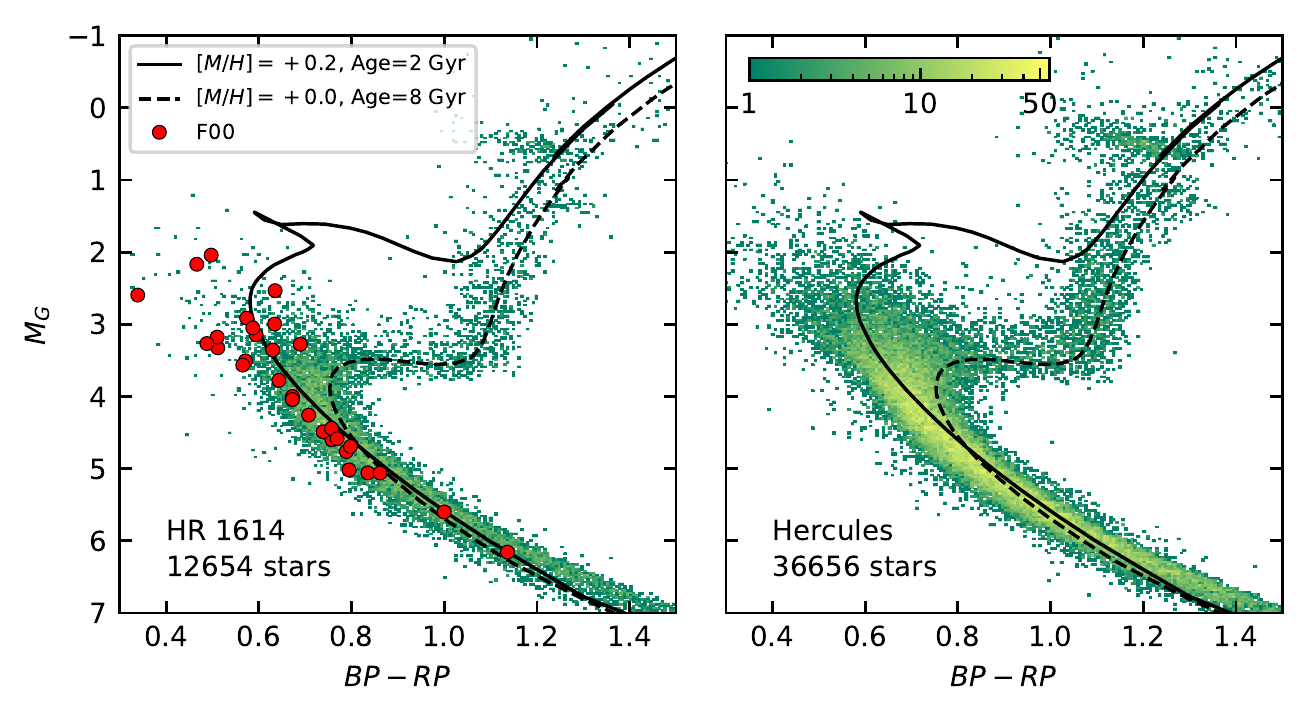}
   }
   \caption{Hertzsprung--Russel diagrams for 12\,654 stars from the HR~1614 moving group (left) and 36\,656 stars from Hercules (right) selected from the black and red ellipses shown in Fig.~\ref{_fig_uv_ellipses}. An isochrone with [M/H] $=+0.2$ and 2 Gyr is shown as a black solid line, and one with [M/H] $=0$ and 8 Gyr is shown as a dashed line. Red dots in the top left plot show all the stars from Table 1 in \citet{_feltzing00}. The colour bar shows the number of stars in each bin for both plots. 
   \label{_fig_hdr_fd}
   }
\end{figure*}

Figure~\ref{_fig_hdr_fd} shows HR diagrams for the stars inside the black and red ellipses that we link to the HR~1614 moving group and the Hercules stream. The red circles show stars from \citet{_feltzing00} that we also could identify in our {\tt StarHorse} sample. Most of these stars follow an $[M/H]=0.2$ and 2 Gyr isochrone, the parameters of HR~1614 that were found by \citet{_feltzing00} and \citet{_desilva07}. At the same time it is clearly seen that stars from ellipses around Hercules and HR~1614 are also composed of multiple stellar populations and cannot be fitted with only one isochrone. From now on we refer to stars inside the black and red ellipses in Fig.~\ref{_fig_uv_ellipses} as the HR~1614 moving group and the Hercules stream, respectively.

\subsubsection{HR diagrams}

\begin{figure*}
   \centering
   \resizebox{0.8\hsize}{!}{
   \includegraphics[viewport = 0  0 640 420,clip]{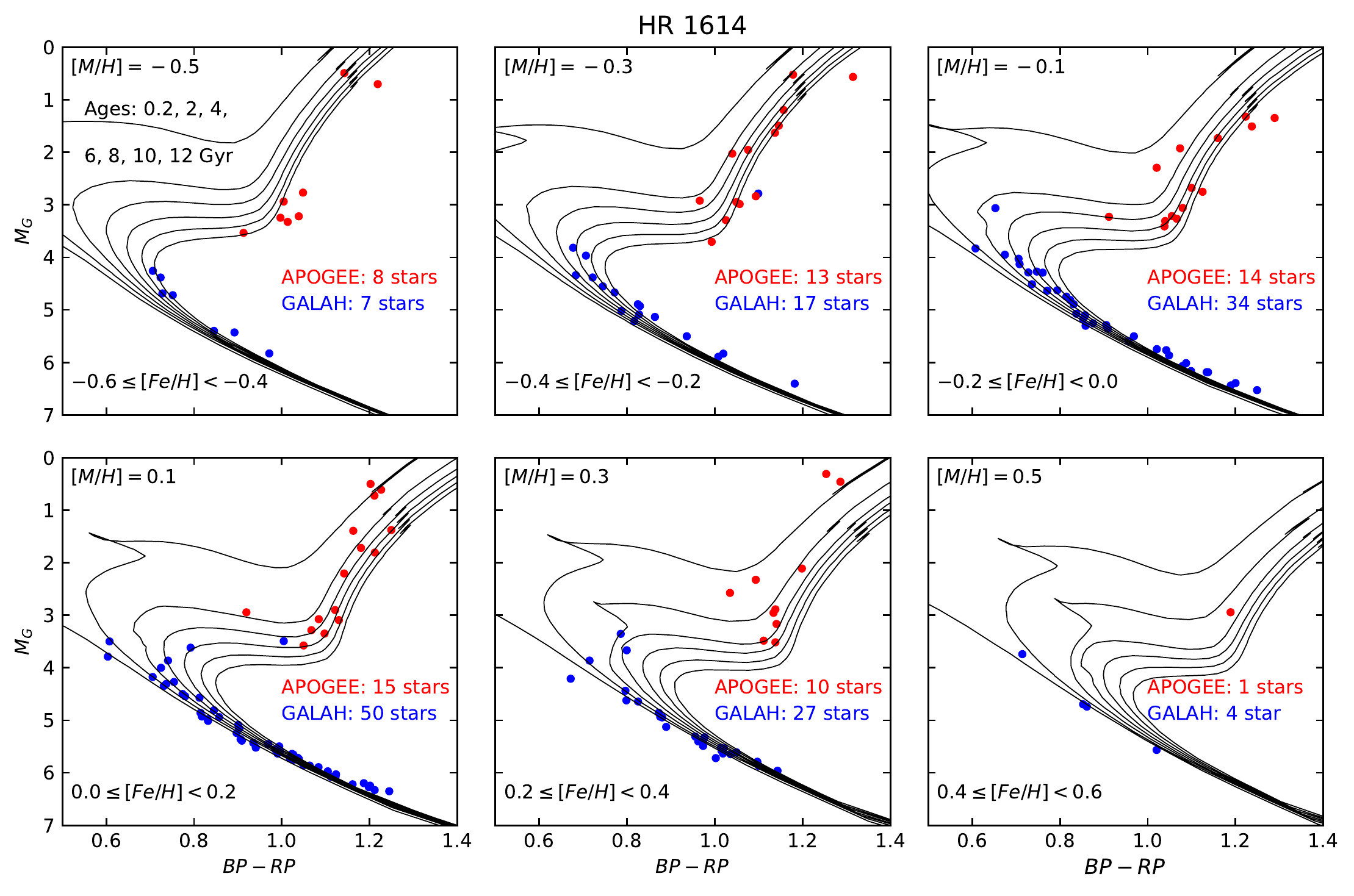}
   }
   \resizebox{0.8\hsize}{!}{
   \includegraphics[viewport = 0  0 640 420,clip]{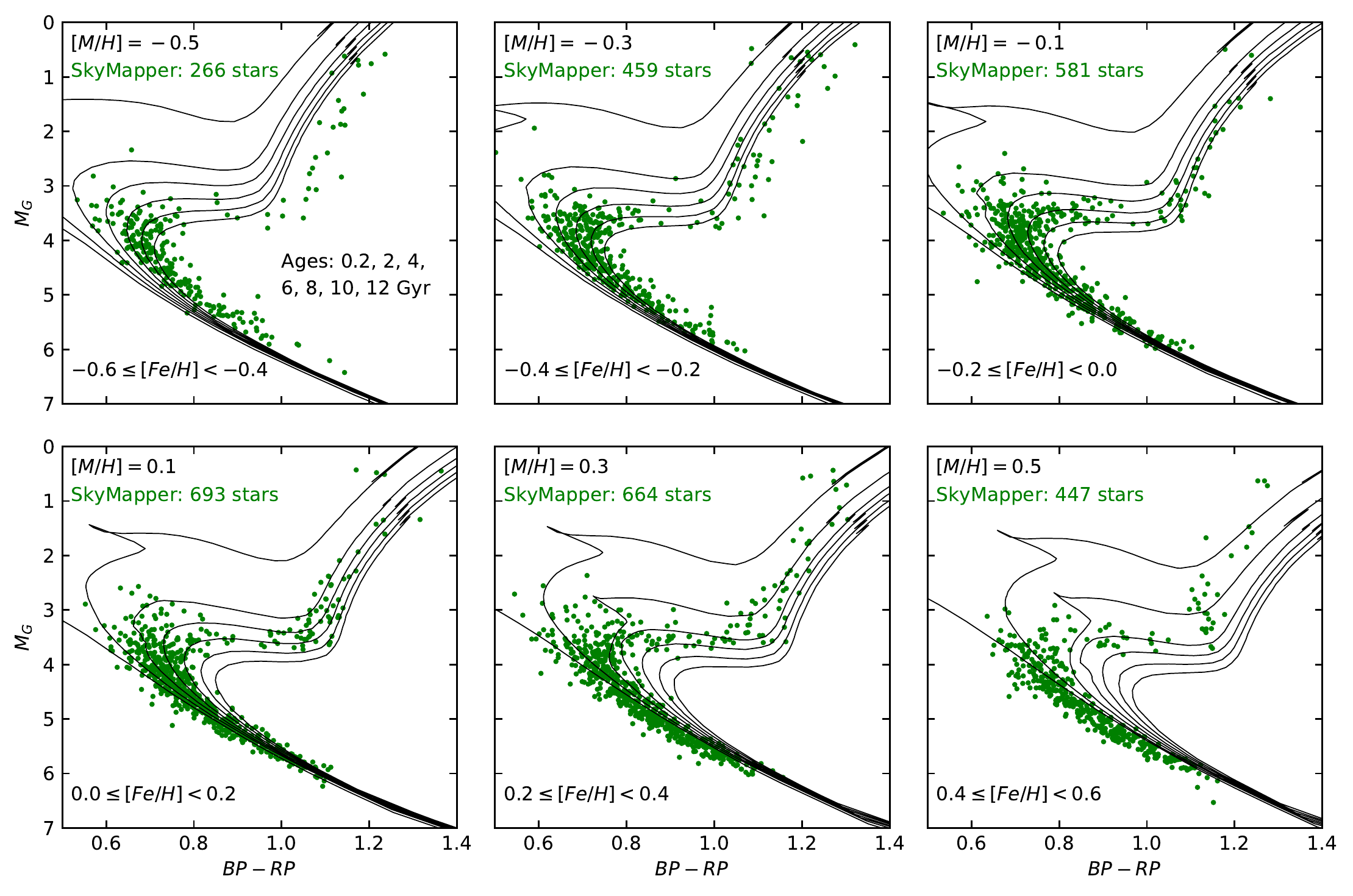}
   }
   \caption{Hertzsprung--Russel diagrams for the stars in the HR~1614 moving group selected from the cross-matched samples between {\tt StarHorse} and APOGEE (red), and {StarHorse} and GALAH (blue), and {\tt StarHorse} and SkyMapper for different metallicity bins. Black lines show isochrones for a mean metallicity of each bin and cover ages 0.2 Gyr and from 2 to 12 Gyr with the steps of 2 Gyr.
   \label{_fig_hrd_hr1614}
   }
\end{figure*}

\begin{figure*}
   \centering
   \resizebox{0.8\hsize}{!}{
   \includegraphics[viewport = 0  0 640 420,clip]{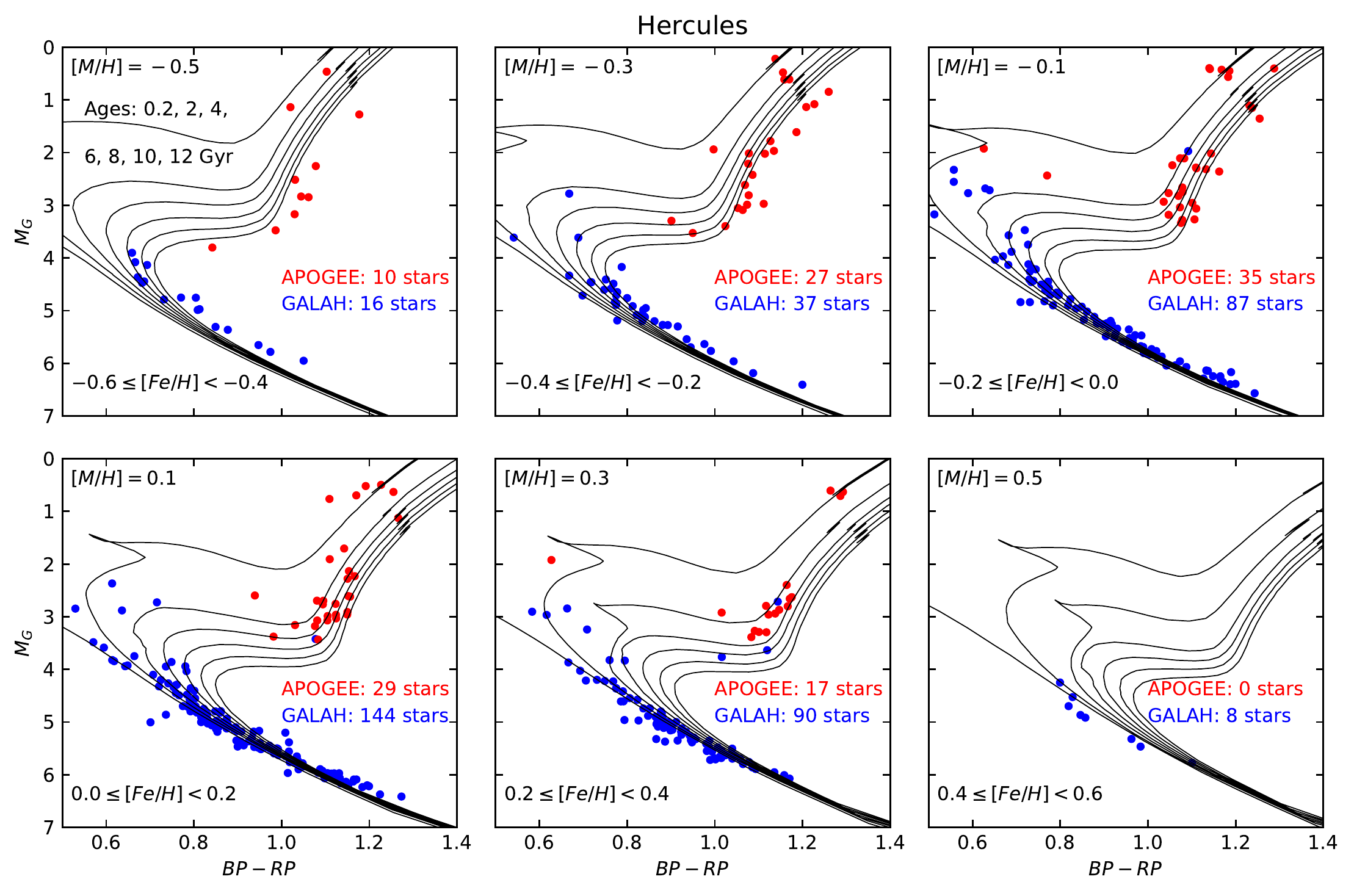}
   }
   \resizebox{0.8\hsize}{!}{
   \includegraphics[viewport = 0  0 640 420,clip]{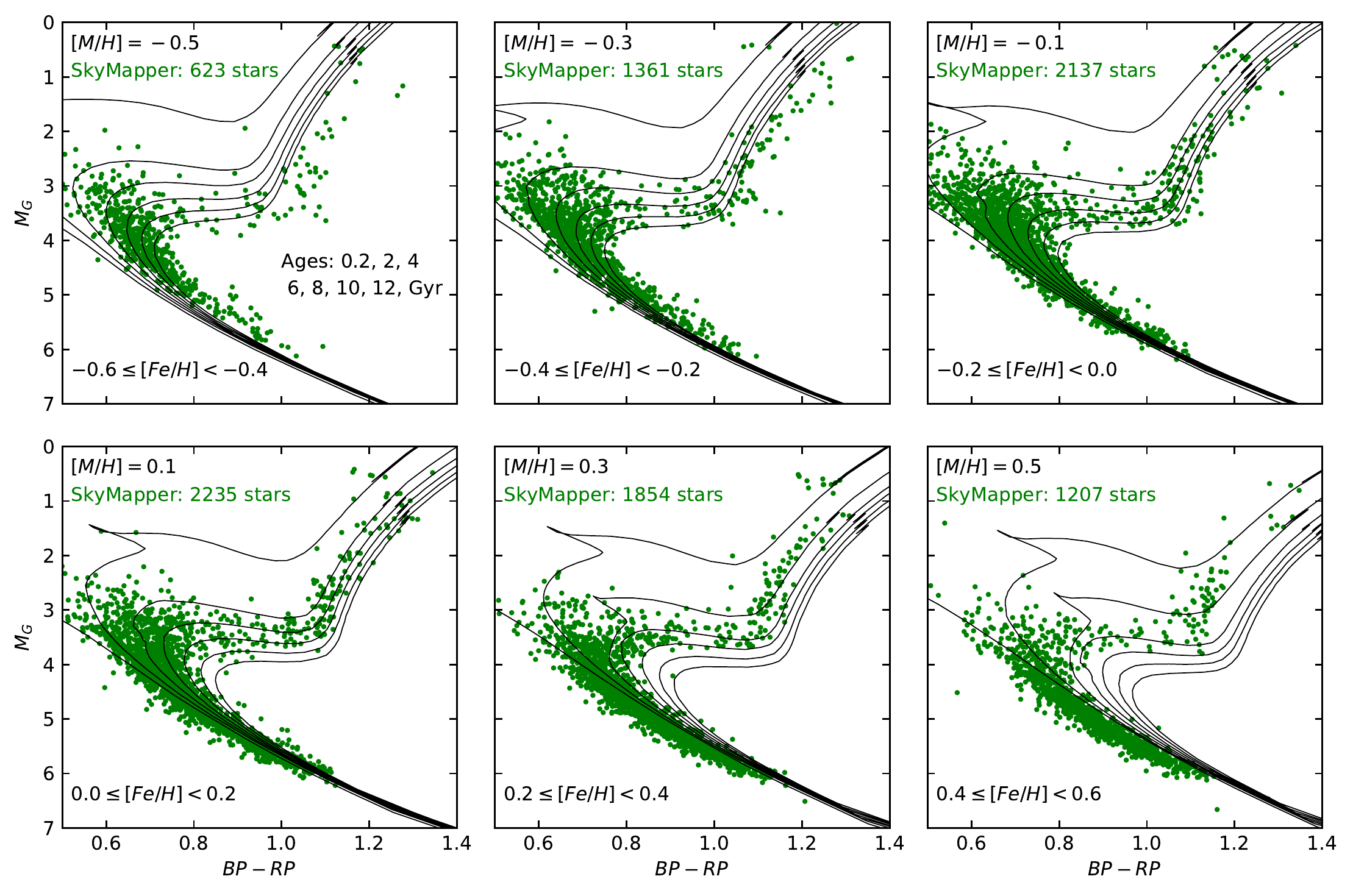}
   }
   \caption{Same as Figure \ref{_fig_hrd_hr1614}, but for the Hercules stream.
   \label{_fig_hrd_herc}
   }
\end{figure*}

\begin{figure*}
   \centering
   \resizebox{0.9\hsize}{!}{
   \includegraphics[viewport = 0  0 430 280,clip]{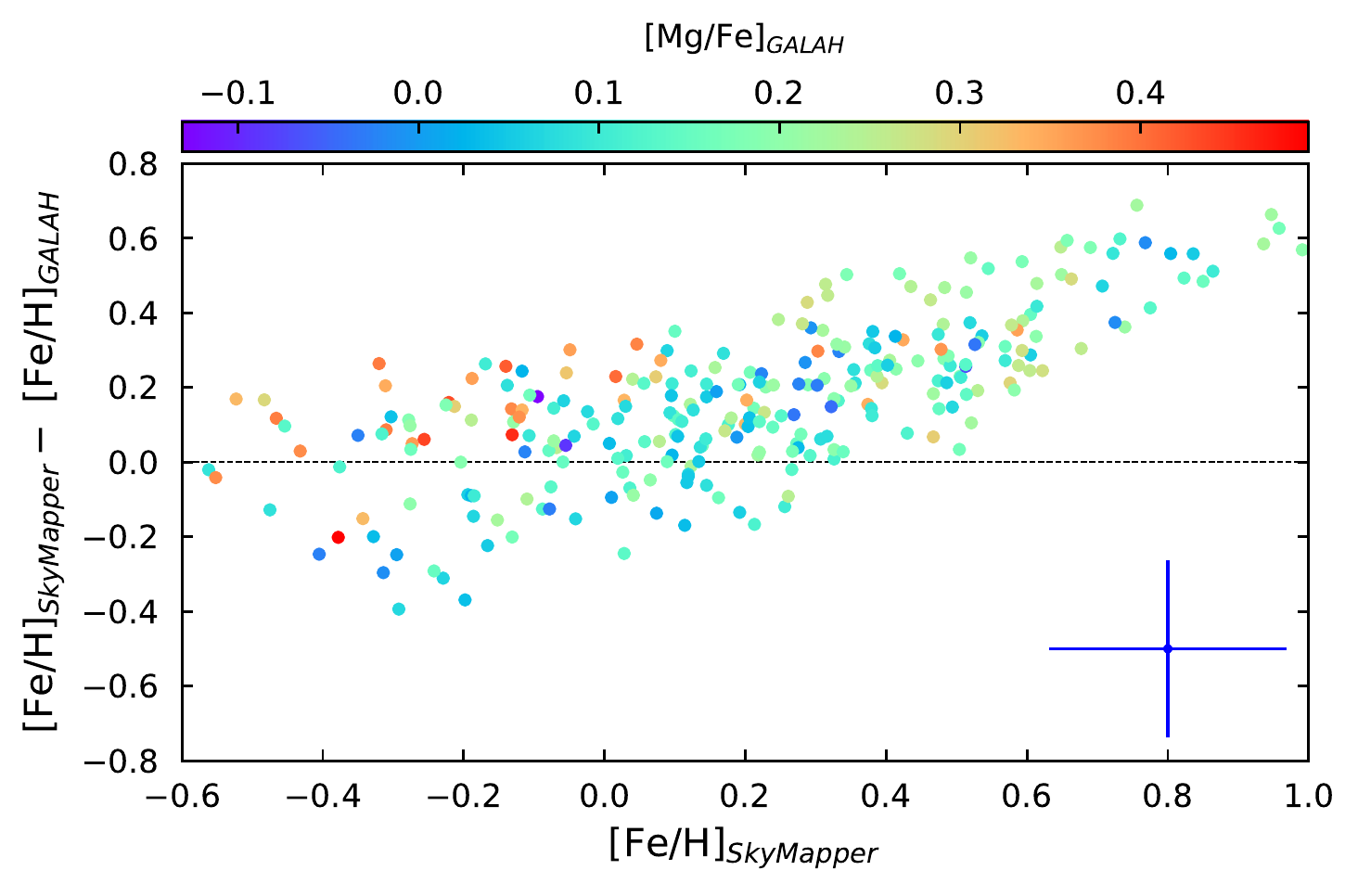}
   \includegraphics[viewport = 0  0 430 280,clip]{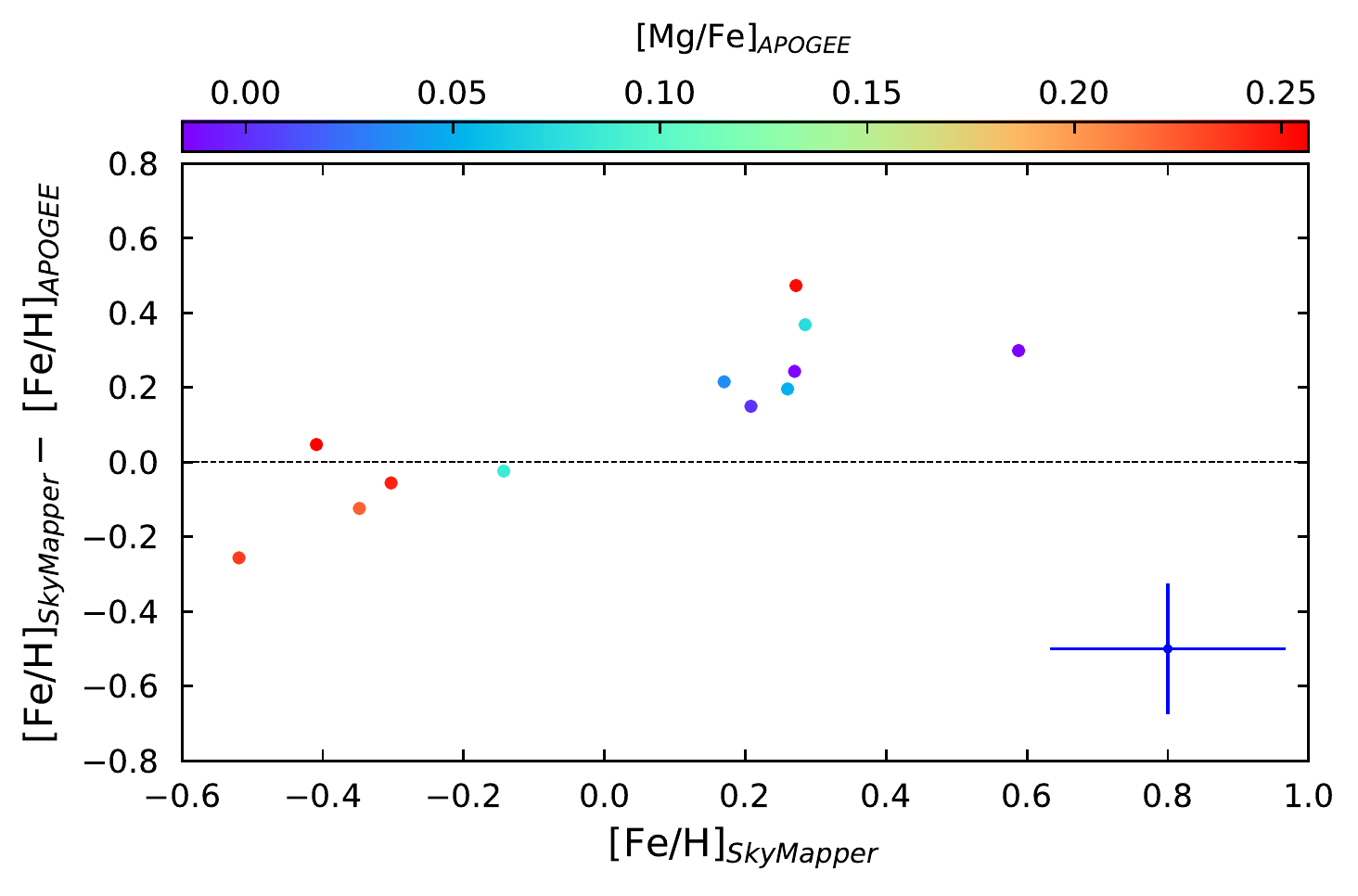}
   }
   \caption{Difference between GALAH and SkyMapper metallicities as a function of SkyMapper metallicity (left). Difference between APOGEE and SkyMapper metallicities as a function of SkyMapper metallicity (right). Both plots are colour-coded by [Mg/Fe] values taken from the GALAH and APOGEE samples, respectively, and are shown for stars in the HR~1614 moving group and the Hercules stream. Blue crosses show typical (mean) error.
   \label{_fig_gal_sm}
   }
\end{figure*}

Figures~\ref{_fig_hrd_hr1614} and \ref{_fig_hrd_herc} show HR diagrams of the HR~1614 moving group and the Hercules stream, respectively, for six metallicity bins in the range $\rm -0.6<[Fe/H]<0.6$. The top panels in each figure show the APOGEE and GALAH data (red and blue dots, respectively), and the bottom panels show stars with SkyMapper data (green dots). We plot isochrones with a metallicity corresponding to the mean metallicity of the stars in each bin. The HR diagrams show that the HR~1614 moving group and the Hercules stream cover a wide range of metallicities and ages. An age--metallicity gradient appears to be present in both groups. Metal-poor stars tend to be older and metal-rich stars seem to be younger, based on a visual inspection of the HR diagrams. It is worth noting that stars in the most metal-poor and metal-rich bins do not perfectly match the isochrones. Therefore, in Fig.~\ref{_fig_gal_sm} we compare metallicities for stars in common between the GALAH and SkyMapper, and the APOGEE and SkyMapper surveys. The difference between SkyMapper and GALAH or APOGEE metallicities increases for metal-rich SkyMapper targets. As the SkyMapper metallicities are estimated based on photometric parameters and calibrated on data from the GALAH survey. The SkyMapper metallicities are overestimated for targets more metal-rich than $\simeq$ 0.5 dex, where the GALAH survey does not provide any stars (see Fig. 11 in \citet{_casagrande19} for more details).  

Overall, the HR diagrams presented in Figs.~\ref{_fig_hdr_fd}, \ref{_fig_hrd_hr1614}, and \ref{_fig_hrd_herc} show that the HR~1614 moving group is a more complex structure than previously thought, and that it does not consist of single-age and single-elemental abundance population of stars, similarly to the Hercules stream. This finding contradicts the hypothesis that the HR~1614 overdensity has a moving group origin, in which case its stars would show homogeneity in age and elemental abundance. 

\subsubsection{Abundance trends and metallicity distribution}
\begin{figure*}
   \centering
   \resizebox{0.9\hsize}{!}{
   \includegraphics[viewport = 0   40 430 280,clip]{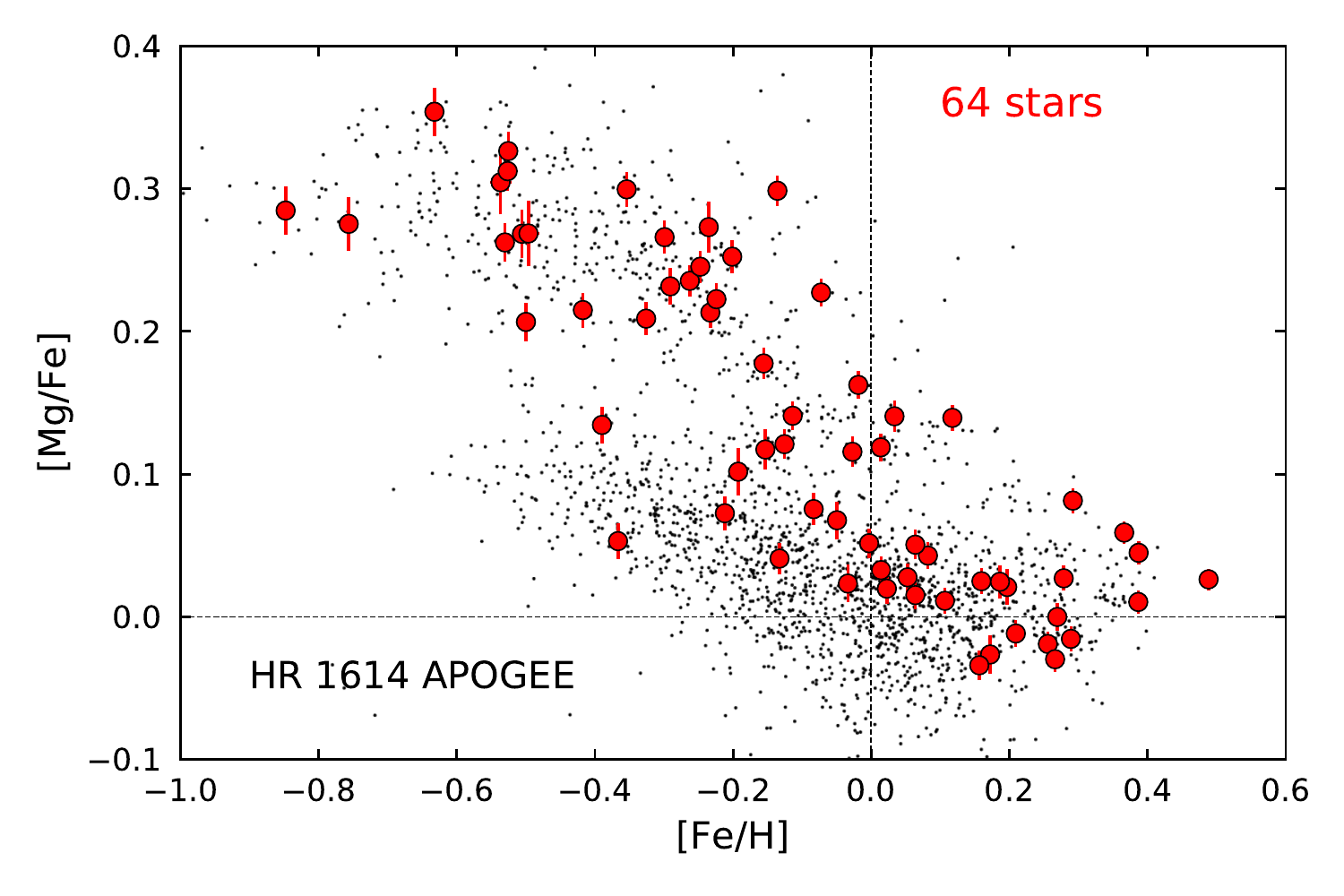}
   \includegraphics[viewport = 25  40 430 280,clip]{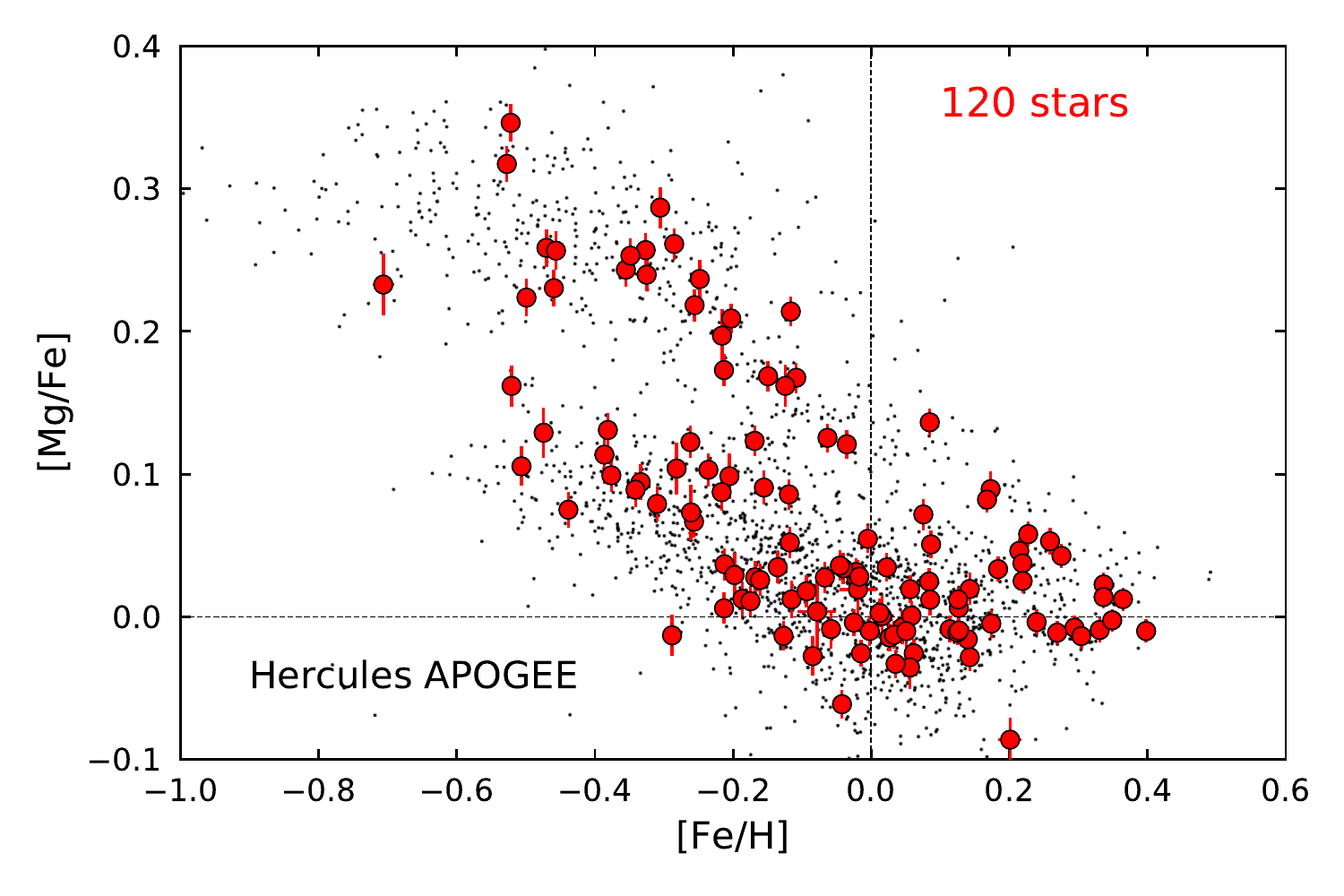}
   }
   \resizebox{0.9\hsize}{!}{
   \includegraphics[viewport = 0   0 430 280,clip]{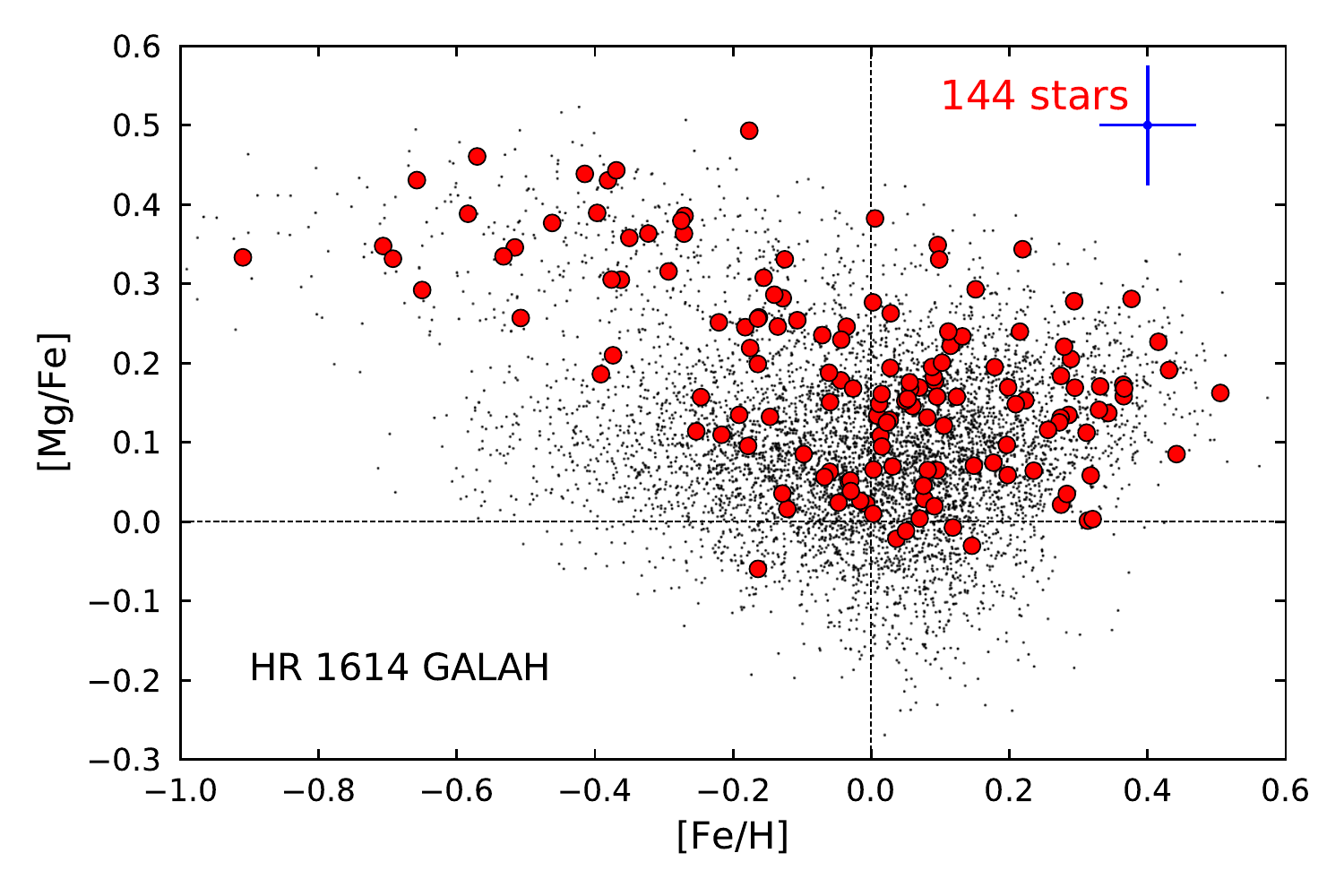}
   \includegraphics[viewport = 25  0 430 280,clip]{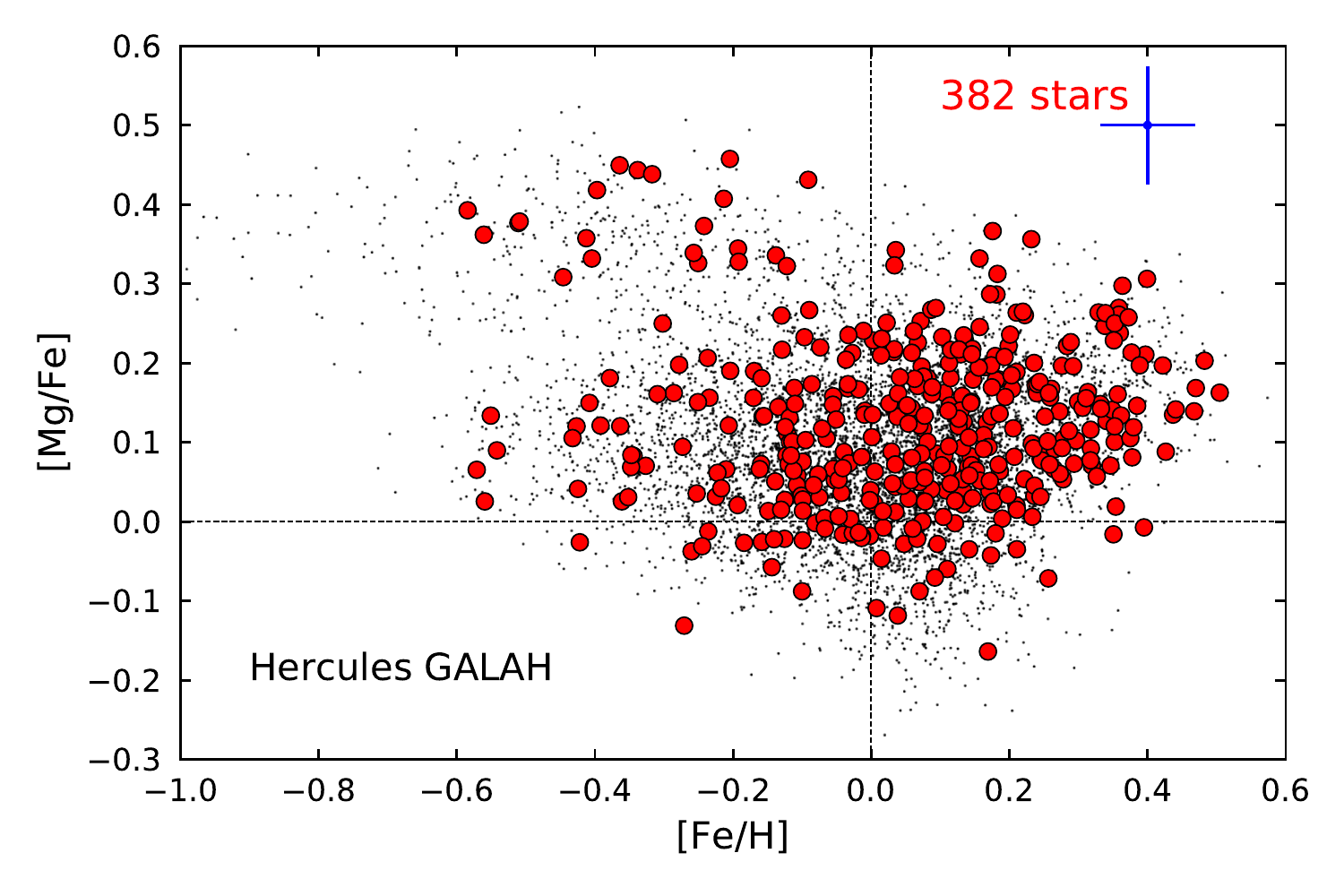}
   }
   \caption{[Mg/Fe]--[Fe/H] diagrams for HR~1614 (left) and Hercules (right) selected from the APOGEE (top) and GALAH (bottom) surveys are shown as red dots. The corresponding uncertainties are shown as error bars for APOGEE distributions and a typical (mean) error is shown for GALAH stars. Background black distributions show stars from APOGEE and GALAH in region 00. 
   \label{_fig_mg_feh}
   }
\end{figure*}

Figure~\ref{_fig_mg_feh} shows the [Mg/Fe] versus [Fe/H] diagrams for stars from the GALAH and APOGEE catalogues that are located in region 00. Candidate members of the HR~1614 moving group (left panels) and the Hercules stream (right panels) are shown as red dots. The GALAH data (lower panels) generally show larger uncertainties than the APOGEE data (upper panels), and therefore we only show a single error bar representing the mean uncertainty for the GALAH stars. Background distributions as well as stars from the groups are composed of two sequences, low- and high-alpha stellar populations, that can be recognised as the chemically defined Galactic thin and thick discs \citep[e.g.][]{_bensby11}. 

\begin{figure*}
   \centering
   \resizebox{0.9\hsize}{!}{
   \includegraphics[viewport = 0   0 400 450,clip]{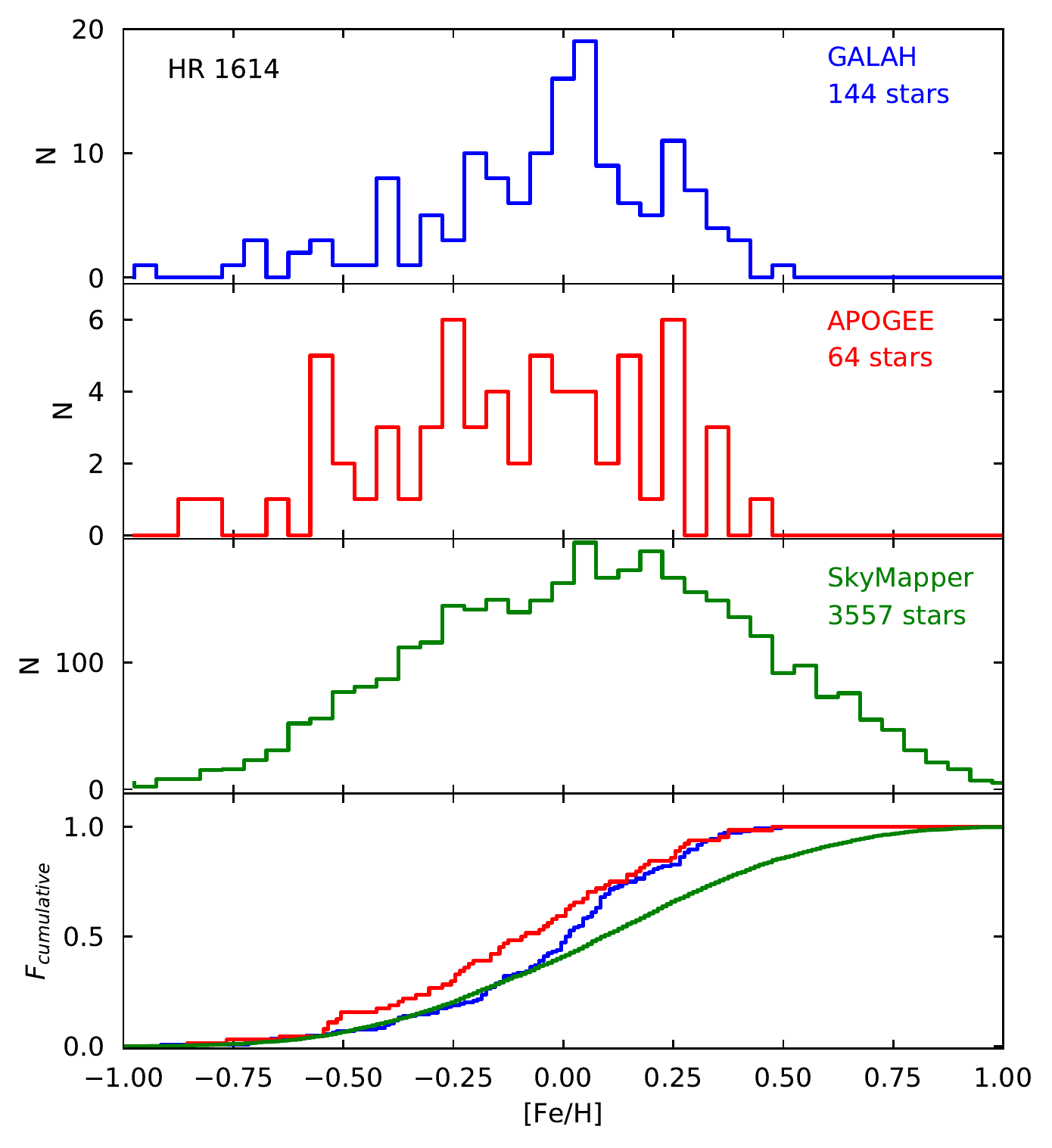}
   \includegraphics[viewport = 0   0 400 450,clip]{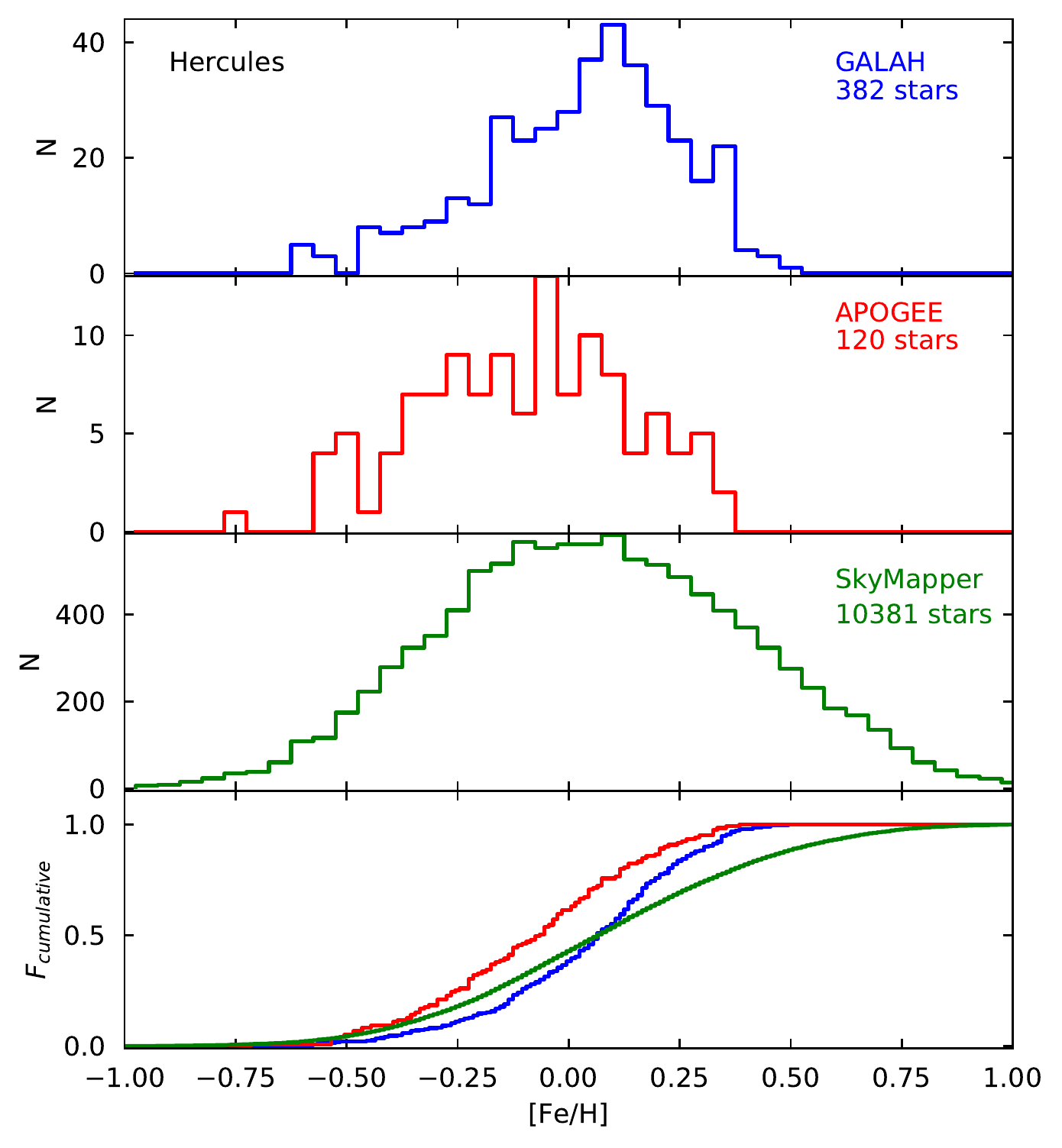}
   }
   \caption{Metallicity distributions for HR~1614 (left) and Hercules (right) from the GALAH (blue), APOGEE (red), and SkyMapper (green) surveys. Both regular histograms (upper three panels) and normalised cumulative histograms (bottom panel) are shown.
   \label{_fig_feh}
   }
\end{figure*}

The metallicity distributions for the HR~1614 moving group and the Hercules stream with GALAH, APOGEE, and SkyMapper data are shown in Fig.~\ref{_fig_feh}. The distributions for the HR~1614 moving group look more spiky due to lower number of stars in the group compared to the Hercules stream. Again, both groups cover a wide range of metallicities and are likely composed of stars that come from different stellar populations. We do not observe the overdensity around [Fe/H] $=0.2$ in the metallicity distribution of the HR~1614 moving group. These findings question the existence of a single metal-rich population of stars reported in \citet{_feltzing00} and \citet{_desilva07} as well as the dissolving cluster origin of the HR~1614 overdensity.

\subsubsection{Age distribution}

\begin{figure*}
   \centering
   \resizebox{0.9\hsize}{!}{
   \includegraphics[viewport = 0   0 1000 500,clip]{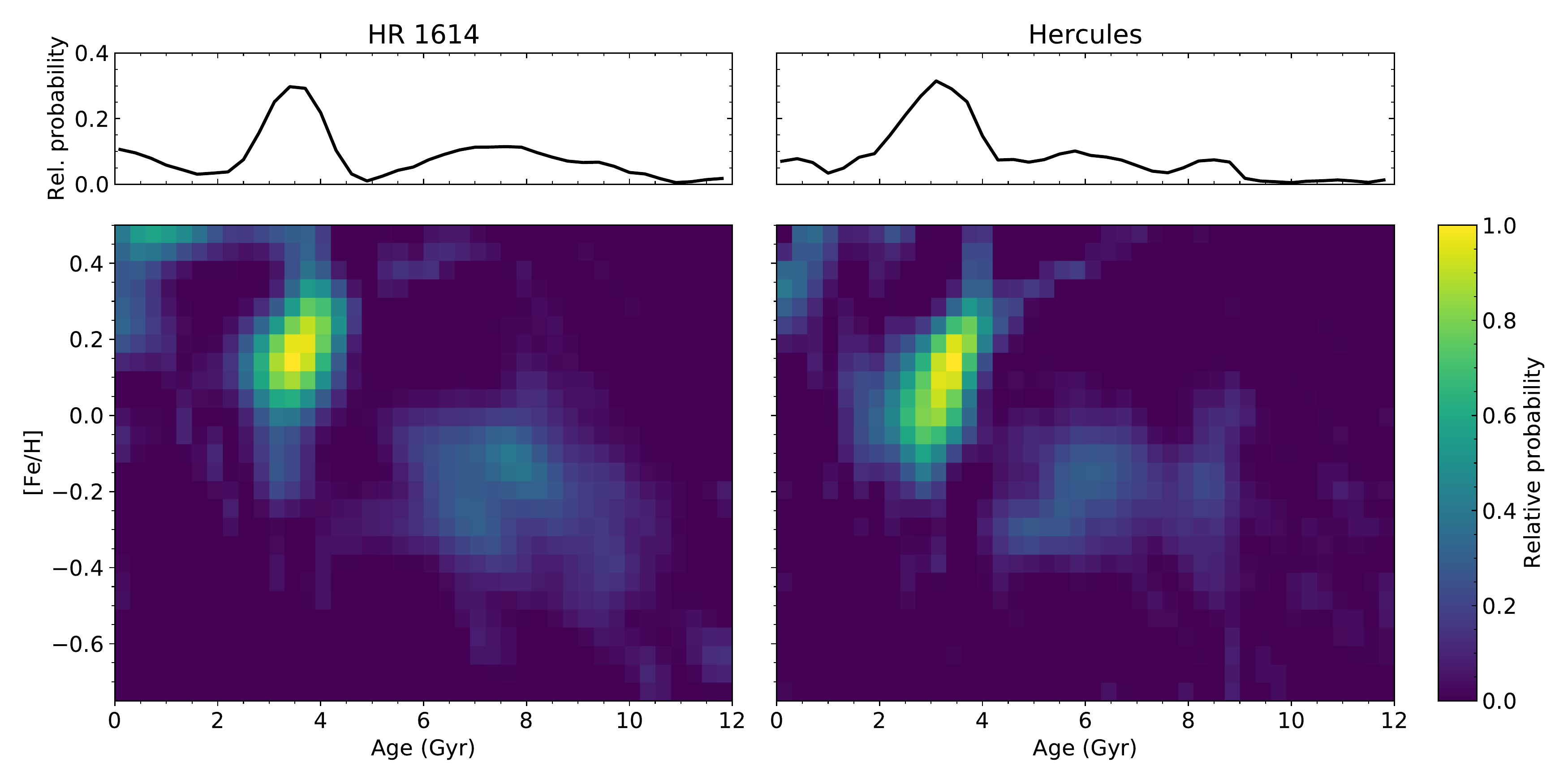}}
   \caption{Age--metallicity probability distributions for the HR~1614 moving group (left) and Hercules stream (right) for stars from the SkyMapper survey. The upper panels show the normalised age distributions after summing over the metallicity dimension.
   \label{_fig_age}
   }
\end{figure*}

Figure~\ref{_fig_age} shows the age--metallicity probability distributions for the HR~1614 moving group and the Hercules streams. To determine these distributions, we first calculated the two-dimensional probability distributions in age and metallicity, $\mathcal{G}_{i}(\tau, [\mathrm{Fe}/\mathrm{H}])$, for each star $i$ following the method by \cite{2019A&A...622A..27H}. In short, the $\mathcal{G}$ function of a star is calculated by fitting a grid of isochrones (PARSEC) to the observed $G$ magnitude, $G_{BP}-G_{RP}$ colour, metallicity, and distance. Then the age--metallicity distribution of the sample as a whole, $\phi(\tau, [\mathrm{Fe}/\mathrm{H}])$, is estimated by maximising the likelihood
\begin{equation}
    L(\phi) = \prod_{i}\int \mathcal{G}_{i}(\tau, [\mathrm{Fe}/\mathrm{H}]) \phi(\tau, [\mathrm{Fe}/\mathrm{H}])\, \mathrm{d}\tau\, \mathrm{d}[\mathrm{Fe}/\mathrm{H}] \, ,
\end{equation}
using an inversion algorithm. The inversion is subject to regularisation which means there is a single free parameter governing the smoothness of $\phi$. Figure~\ref{_fig_age} shows the results for one choice of this parameter, but we have tested a wide range of values and find that the results are similar enough to not affect our conclusions. This method will be described in detail in an upcoming publication (Sahlholdt \& Lindegren, in prep).

Both groups are composed of at least two stellar populations. The first population is metal-poor and old with the centre at $\rm [Fe/H]\approx-0.2$ and an age of about 8\,Gyr. The second population is metal-rich and young with the centre at $\rm [Fe/H]\approx+0.1$ and an age of about 3\,Gyr. These two populations can most likely be associated with the Galactic thin and thick discs. There is also a third clump at a very young age, less than about 1\,Gyr. These are stars with ages that are located at the edge of the grid and are artefacts of the age estimation procedure.

From this analysis it is clear that the HR~1614 overdensity is not a 2\,Gyr old stellar population, but rather is composed of a mix of stars from the Galactic disc. The presence of an age bi-modality in the HR~1614 overdensity also contradicts the idea of its dissolving cluster origin.

\section{Origin of the HR~1614 overdensity}\label{_sec_discussion}

\subsection{Dissolving open cluster origin}
Stars that originate from a dissolving open cluster were formed from the same molecular cloud, and thus, retain similar chemical compositions and ages \citep[]{_freeman02}. As was proposed in \citet{_eggen71, _eggen92, _eggen98}, for a long time HR~1614 was considered a classical moving group. The analysis of Hipparcos stars accompanied by photometric and spectroscopic data by \citet{_feltzing00} and \citet{_desilva07} showed that HR~1614 was a unique stellar population with a metallicity $\rm [Fe/H]\approx 0.2$, and age of about 2\,Gyr with a very small scatter in abundances in other elements. However, we do not see any evidence that the HR~1614 overdensity is a mono-age and mono-abundance stellar population in our study. The analysis of HR diagrams, metallicity, and age distributions presented in Sect.~\ref{_sec_analysis} clearly shows that HR~1614 is a mix of two stellar populations that resemble properties of the Galactic thin and thick discs. A small scatter in abundances observed in \citet{_desilva07} is, possibly, a result of the stellar sample analysed in their work. It is mainly composed of stars from \citet{_feltzing00}, who selected metal-rich stars from their Box 5 as possible members of the HR 1614 moving group. This selection effect can potentially explain the small scatter in abundances reported in \citet{_desilva07}.

\begin{figure*}
   \centering
   \resizebox{0.9\hsize}{!}{
   \includegraphics[viewport = 0  0 410 280,clip]{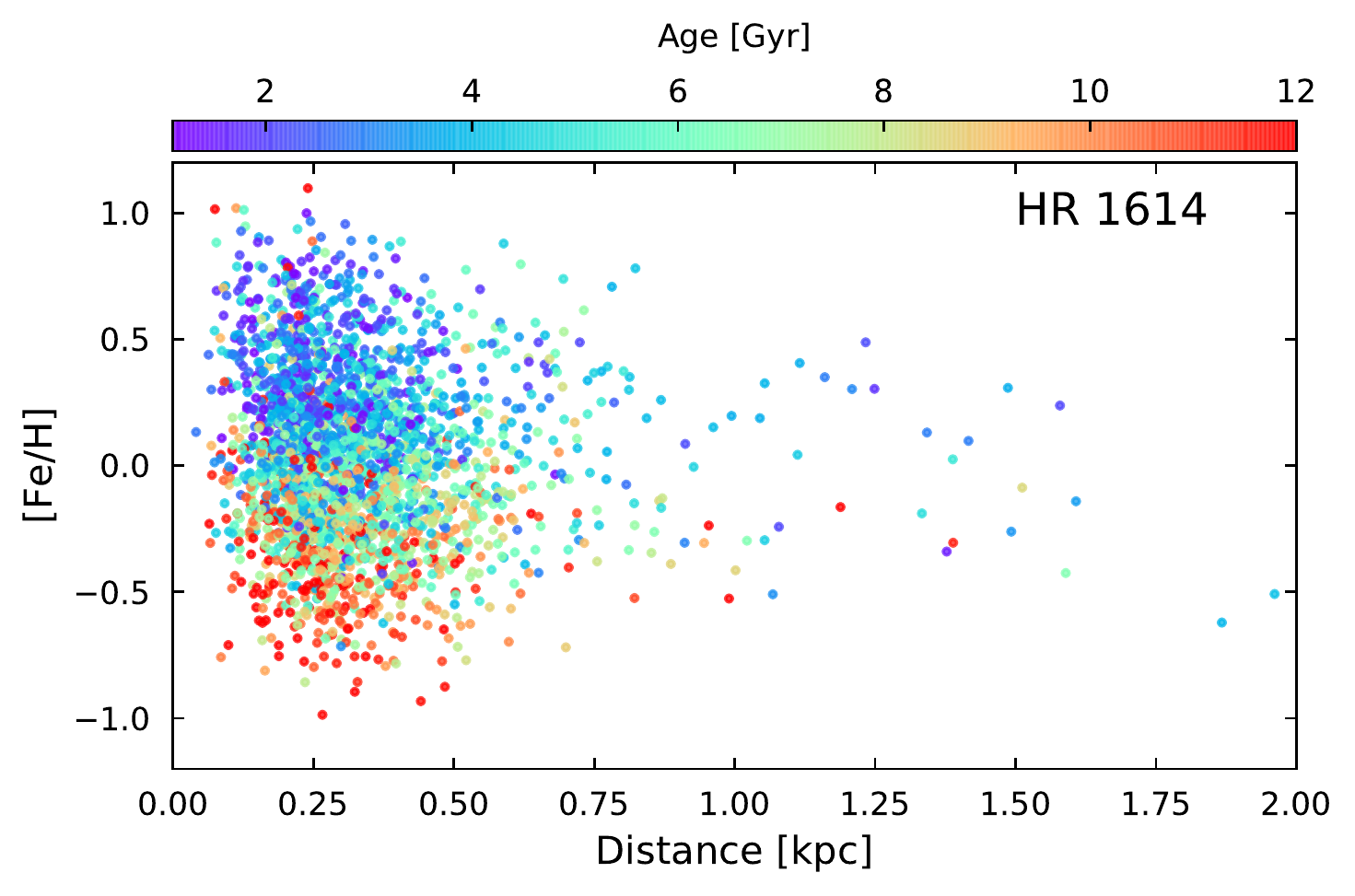}
   \includegraphics[viewport = 0  0 410 280,clip]{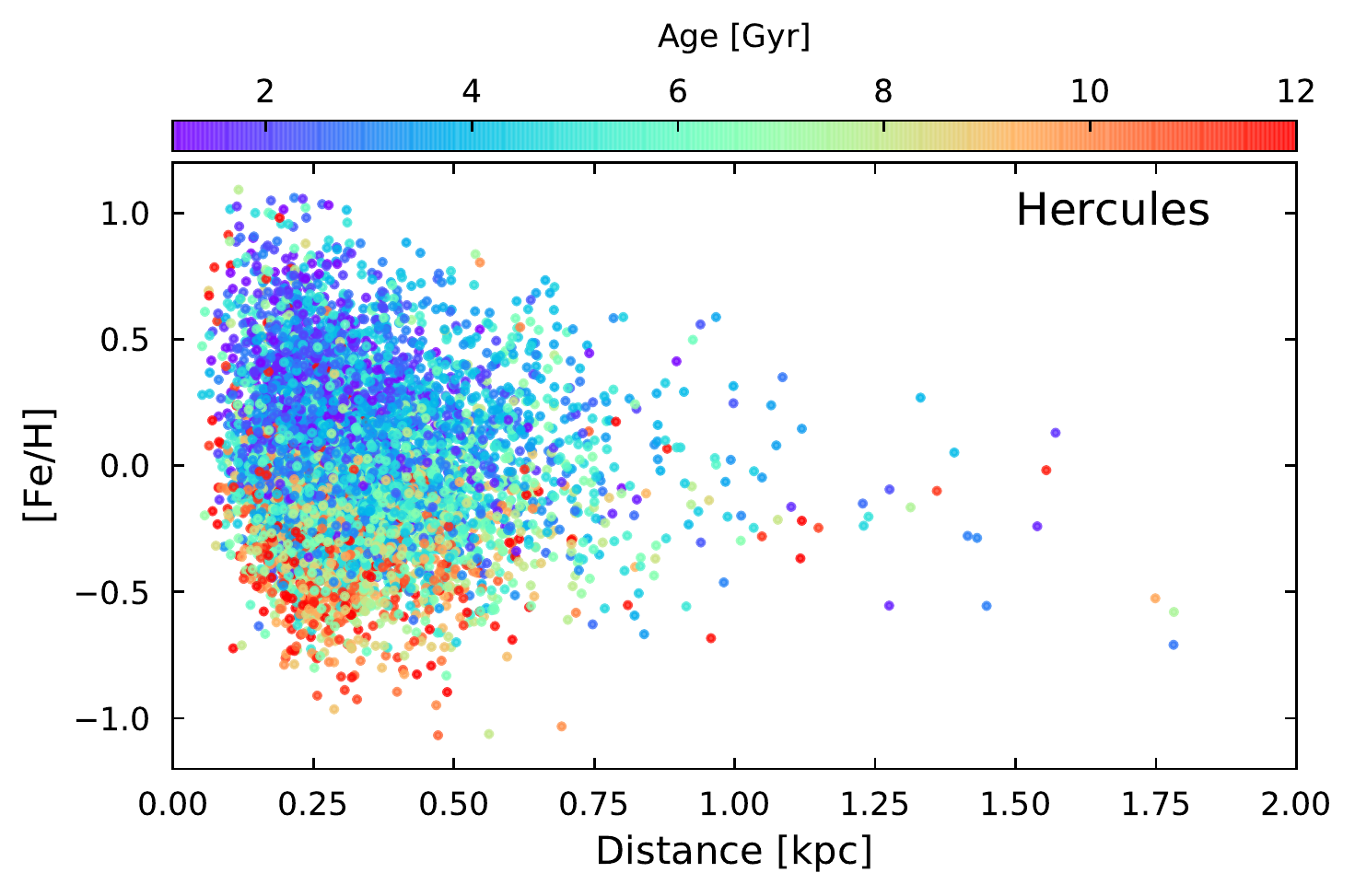}
   }
   \caption{Distance relative to the Sun vs. metallicity diagram colour-cored by age for SkyMapper stars in the HR~1614 moving group (left) and the Hercules stream (right). 
   \label{_fig_age_dist}
   }
\end{figure*}

In this work we selected 12\,654 targets in the HR~1614 moving group from our data sample. This is a significantly larger number of stars-members of the group than ever analysed before. Pre-{\it Gaia} works studied stars that are located roughly within 100 pc around the Sun and found that the HR~1614 overdensity is a single age and abundance population \citep[e.g.][]{_feltzing00, _desilva07}. Since our stellar sample covers a larger volume around the Sun, we check if young stars with higher than solar metallicities are located closer to the Sun in Fig.~\ref{_fig_age_dist}. We do not observe any peculiarities in distributions shown for SkyMapper stars in Fig.~\ref{_fig_age_dist}. Both distributions look similar to younger stars in the metal-rich part of the diagram and older stars in the metal-poor part. It also shows that most of the stars are located within a 500\,pc radius around the Sun. There is no unique, young, and nearby stellar population with higher-than-solar metallicity neither in the HR~1614 nor in the Hercules velocity overdensities. This means that HR~1614 should not be considered a dissolving open cluster any longer.     

\subsection{Accretion origin}
Stars that originate from a dwarf galaxy or satellite and were accreted to the Milky Way retain unique chemical composition, ages, and velocities that are usually different from the background Milky Way stars \citep[e.g.][]{_tolstoy09, _ruchti15, _helmi20}. Recent examples of such populations are Gaia-Sequoia \citep[e.g.][]{_myeong19}, Gaia-Sausage/Gaia-Enceladus \citep[e.g.][]{_belokurov18, _helmi18}. Based on the metallicity, velocity, and age distributions discussed in Sect.~\ref{_sec_analysis}, it is clear that the HR~1614 and Hercules overdensities are not accreted stellar populations. The dynamical and chemical properties of the groups are similar to the vast majority of the Milky Way disc stars. It is unlikely that HR~1614 is an accreted stellar population. 

\subsection{Resonant origin}
Stars that were influenced by resonances with the bar and/or spiral arms can be observed as kinematic structures that consist of mixed stellar populations. Hercules is a group that is widely discussed in the context of resonances. As was first proposed in \citet{_dehnen00}, the Hercules stream is possibly caused by the outer Lindblad resonance (OLR) of the Galactic bar. Later it was proposed that the Hercules stream is likely a result of a corotation resonance (CR) with the bar \citep[e.g.][]{_perez17, _binney19}. Recent results from \citet{_hunt19} show that it is possible to reproduce the Hercules stream in simulations by combining bar and multiple spiral structures. In their Figure 2 \citet{_hunt19} call the bottom-most structure a part of the Hercules stream. If we convert $V$ velocity of the HR 1614 group to the vertical component of the angular momentum $L_z$, assuming that the Sun is located at 8.34 kpc from the Galactic centre, we get $\sim1340$ kpc $\kms$. This is similar to the value found in \citet{_hunt19}. \citet{_hunt19} and we use the same values that we do for the peculiar motion of the Sun. It is exactly the same position in the distribution where the HR~1614 overdensity is supposed to be located, and thus, should be called the HR~1614 moving group instead. 

Many studies show that the Hercules stream conserves a vertical component of angular momentum with Galactocentric radii which is a signature of a resonant origin \citep[e.g.][]{_ramos18, _hunt19, _kushniruk19}. In Figs.~\ref{_fig_uv_ellipses} and \ref{_fig_uv} we clearly see that the Hercules stream is not tilted in the $U-V$ space. It covers a wide range of negative $U$ velocities, but its $V$ velocity is $-50 \kms$ and is almost constant. This means that the vertical component of angular momentum $L_z$ is conserved (taking into account that $V$ is proportional to $L_z$). Unlike the Hercules stream, the HR~1614 overdensity is tilted in the $U-V$ space. This tilt was explained in \citet{_feltzing00} in dynamical simulations of a dissolving open cluster. At the same time \citet{_feltzing00} found that the HR~1614 overdensity is homogeneous in age and metallicity, which we do not observe in our work. The HR~1614 overdensity is also present in other four regions shown in Figure \ref{_fig_uv}. It almost retain $L_z$ with $R$. Taking into account that the HR~1614 overdensity is mainly located at negative $U$ and its proximity to the Hercules stream, the resonant origin of the HR~1614 overdensity should not be excluded.

\subsection{Phase mixing origin}
The concept of phase mixing implies that a phase-space (position and momentum variables) full of overdensities will evolve with time to a stationary state \citep{_tremaine99}. A moving groups is one example of a phase mixing process. In the recent literature, phase mixing is mainly discussed in the context of internal or external perturbation mechanisms that can induce the process on large scales \citep[e.g.][]{_laporte19, _hunt19}. This discussion is especially pertinent after the discovery of a phase spiral by \citet{_antoja18}, which is strong evidence of an ongoing phase mixing in the Galactic disc. 

Symmetric arches of constant energy around $U=0 \kms$ with 20 $\kms$ separation in $V$ were predicted in the model that was first proposed by \citet{_minchev09}, where a perturbation with a dwarf galaxy, that took place roughly 2\,Gyr ago, leads to a phase wrapping. This external perturbation event could have contributed to a velocity distribution and can be observed as overdensities aligned across lines of constant energy. \citet{_minchev09} discuss the HR~1614 overdensity in the context of a dissolving cluster and try to fit it into their ringing theory. They state that the HR~1614 moving group is elongated in $U$ and curved in $V$ in their Figure 3, which makes is consistent with a wrapping origin. The fact that the moving group is not symmetric and is elongated towards negative $U$ is explained by \citet{_minchev09} as being due to the proximity to the Hercules stream, which is most likely caused by the Galactic bar. In summary, \citet{_minchev09} suggest that the HR~1614 overdensity is a dissolving open cluster that formed earlier than an external perturbation event took place and was distorted by phase wrapping, forming almost an arch of constant energy, and by resonances with the bar, causing elongation towards negative $U$. 

Since the arches predicted by \citet{_minchev09} are symmetrical and the observed velocity distribution has non-asymmetries, \citet{_quillen18_galah} proposed another explanation, where asymmetric arches are formed due to stars crossing spiral arms. At the same time their theory cannot explain vertical velocity distribution, and thus external perturbation is not excluded. \citet{_hunt19} explored various combinations of spirals and bar resonances, and stars crossing dissolving transient spirals and found that there are many models that can successfully explain the main velocity overdensities. Some simulations reproduce a velocity overdensity at the place where the HR~1614 moving group is supposed to be. \citet{_khanna19} showed that phase mixing of disrupting spiral arms and phase mixing due to external perturbation can generate arches and ridges, but they do not talk specifically about the HR~1614 moving group in their work. 

In our work we clearly see that the HR~1614 overdensity is not a dissolving open cluster due to chemical and age spread in the group of stars. We also observe that the HR~1614 overdensity is located mainly at negative $U$. At the same time the group is tilted in $V$, but does not form a complete arch of constant energy. This points towards a complex origin of the HR~1614 overdensity. The stars could be influenced by transiting spirals and bar that would shift it towards negative $U$. Phase mixing due to disrupting spiral structure or ringing due to external perturbation could have contributed to the tilt in $V$.    

\section{Conclusions}\label{_sec_conclusion}
In this paper we revised the origin of the HR~1614 moving group. We analysed a combination of {\it Gaia} and {\tt StarHorse} catalogues and investigated kinematic and photometric properties of stars in a small volume near to the Sun. To study chemical properties of the group we accompanied kinematic data with elemental abundances from the APOGEE and GALAH surveys and photometric metallicities SkyMapper survey. For SkyMapper stars Bayesian ages were calculated for stellar populations. Combining these surveys allowed us to analyse a significantly larger data set with higher precision in the astrometric, photometric, and spectroscopic measurements compared to previous studies. Using the methods developed in \citet{_kushniruk19} we have improved the selection of potential members. These two advances allowed us to better explore the possibility that HR 1614 is or is not a dissolving open cluster. We also compared the properties of the HR 1614 moving group with the Hercules stream which is the nearest velocity overdensity to the HR 1614 moving group. The Hercules stream is most likely of resonant origin, making it a valuable comparison to determine whether HR1614 is a single stellar population. The main results of the paper are the following:  

\begin{itemize}
    \item The HR~1614 overdensity is clearly present in the $U-V$ velocity distribution at $U\simeq-20$, $V\simeq-60 \kms$. This location is consistent with the results from previous studies  \citep[e.g.][]{_eggen98, _feltzing00, _ramos18, _kushniruk19}. 
    \item The HR diagrams, metallicity, and age distributions show that HR~1614 consists of two stellar populations that have properties similar to the thin and thick discs. 
    \item The HR~1614 overdensity is mainly present at negative $U$ velocities, which could point towards a resonant origin similarly to the Hercules stream. At the same time the HR~1614 overdensity is slightly tilted in $V$, which means that the group has a variation in its vertical angular momentum distribution.  
    \item The HR~1614 overdensity does not form a complete arch of constant energy in the $U-V$ space, which would be an indication of a phase mixing origin. 
\end{itemize}    

Based on the above, it is clear that the HR~1614 overdensity is neither a dissolving open cluster nor an accreted stellar population. We conclude that the HR~1614 overdensity has a complex origin which is a combination of various dynamical mechanisms. As in the case of the Hercules stream, the HR~1614 overdensity could have formed due to different types of resonances such as the CR or OLR with the Galactic bar. This together with phase mixing due to disrupting spiral arms could explain the shape and location of the group in the $U-V$ space. Phase mixing due to external perturbation with a dwarf galaxy or a satellite that happened 2\,Gyr ago, as proposed in \citet{_minchev09}, is also possible. The mixed populations that we observe in the HR~1614 overdensity and the Hercules stream are older than a potential perturbation event. Also in \citet{_kushniruk19} we observed kinematic structures with roughly 20 $\kms$ separation in $V$, especially at lower $V$, which is consistent with a ringing event proposed in \citet{_minchev09}. 

In order to disentangle how many different dynamical mechanisms have actually contributed to the formation of velocity structures like the HR~1614 overdensity, further studies are required. Numerical simulations combined with detailed investigation of elemental abundances from spectroscopic surveys like WEAVE \citep{dalton2014}, 4MOST \citep{_dejong2019}, and the Gaia-ESO survey \citep{_gilmore12}, and kinematics from upcoming Gaia data releases will provide more information about the origin of kinematic structures.

\begin{acknowledgements}
T.B. was funded by grant No. 2018-04857 from the Swedish Research Council. L.C. acknowledges support from the Australian Research Council Future Fellowship FT160100402. C.L.S., D.F., and S.F. were supported by the project grant The New Milky Way from the Knut and Alice Wallenberg foundation and by the grant 2016-03412 from the Swedish Research Council. 
\end{acknowledgements}


\bibliographystyle{aa}
\bibliography{references}
\end{document}